\documentclass[10pt]{article}

\usepackage{pstricks}
\usepackage{pst-plot}
\usepackage{pst-node}
\usepackage{amsmath}
\usepackage{pst-grad}
\usepackage{amssymb}
\usepackage{xspace}
\usepackage{pst-math}
\usepackage{pst-xkey}
\usepackage{pst-coil}
\usepackage{pst-3dplot}
\usepackage[T1]{fontenc}
\usepackage{ae}
\usepackage[ansinew]{inputenc}

\newrgbcolor{darkred}{0.8 0 0}
\newrgbcolor{darkerred}{0.7 0 0}
\newrgbcolor{darkestred}{0.5 0 0}
\newrgbcolor{darkgreen}{0 0.5 0.5}
\newrgbcolor{darkergreen}{0 0.6 0}
\newrgbcolor{darkestgreen}{0 0.5 0}
\newrgbcolor{darkblue}{0 0 0.6}
\newrgbcolor{darkestblue}{0 0 0.5}
\newrgbcolor{redgreen}{0.8 0.8 0}
\newrgbcolor{bluegreen}{0 0.4 0.7}
\newrgbcolor{bluewhite}{0 0.9 1}
\newrgbcolor{lightgreen}{0.6 1 0.6}
\newrgbcolor{lightred}{1 0.8 0.8}
\newrgbcolor{lightyred}{1 0.4 0.4}
\newrgbcolor{lightyblue}{0.5 0.5 1}
\newrgbcolor{lightblue}{0.8 1 1}
\newrgbcolor{lilas}{0.5 0.3 0.7}
\newrgbcolor{lilaswhite}{1 0.8 1}
\newrgbcolor{redy}{1 0.6 0.6}
\newrgbcolor{bluish}{0.6 0.6 1}
\newrgbcolor{sand}{0.9 0.9 0.7}
\newrgbcolor{brown}{0.4 0.4 0}
\newrgbcolor{blue1}{0 0.35 0.65}
\newrgbcolor{blue2}{0 0.4 0.6}
\newrgbcolor{blue3}{0 0.45 0.55}
\newrgbcolor{blue4}{0 0.5 0.5}
\newrgbcolor{blue5}{0 0.55 0.45}
\newrgbcolor{blue6}{0 0.6 0.4}
\newrgbcolor{blue7}{0 0.65 0.35}

\newcommand{\R }{\mathbb R}
\newcommand{\N }{\mathbb N}
\newcommand{\Z }{\mathbb Z}
\newcommand{\Q }{\mathbb Q}
\newcommand{\C }{\mathbb C}
\newcommand{\sen }{\ \! {\rm sen}\ \! }
\newcommand{\tg }{\ \! {\rm tg}\ \! }
\newcommand{\cosec }{\ \! {\rm cosec}\ \! }
\newcommand{\cotg }{\ \! {\rm cotg}\ \! }
\newcommand{\dx }{\ \! dx}
\newcommand{\e }{\ \! e}
\newcommand{\senh }{\ \! {\rm senh}\ \! }
\newcommand{\tgh }{\ \! {\rm tgh}\ \! }
\newcommand{\cotgh }{\ \! {\rm cotgh}\ \! }
\newcommand{\sech }{\ \! {\rm sech}\ \! }
\newcommand{\cosech }{\ \! {\rm cosech}\ \! }
\newcommand{\arcsen }{\ \! {\rm arcsen}\ \! }
\newcommand{\arctg }{\ \! {\rm arctg}\ \! }
\newcommand{\arccotg }{\ \! {\rm arccotg}\ \! }
\newcommand{\arcsenh }{\ \! {\rm arcsenh}\ \! }
\newcommand{\arccosh }{\ \! {\rm arccosh}\ \! }
\newcommand{\arctgh }{\ \! {\rm arctgh}\ \! }
\newcommand{\arccosech }{\ \! {\rm arccosech}\ \! }
\newcommand{\arcsech }{\ \! {\rm arcsech}\ \! }
\newcommand{\arcsec }{\ \! {\rm arcsec}\ \! }
\newcommand{\arccosec }{\ \! {\rm arccosec}\ \! }
\renewcommand{\arctgh }{\ \! {\rm arctgh}\ \! }
\newcommand{\arccotgh }{\ \! {\rm arccotgh}\ \! }
\newcommand{\nega }{\neg \ }
\newcommand{\eq }{\Leftrightarrow }

\begin{document}

\title{Shocks in financial markets, price expectation, and damped harmonic oscillators}

\author{Leonidas Sandoval Junior\thanks{E-mail: leonidassj@insper.edu.br (corresponding author)} \\ Italo De Paula Franca \thanks{E-mail: italopf@al.insper.edu.br} \\ \\ Insper, Instituto de Ensino e Pesquisa}

\maketitle

\begin{abstract}
Using a modified model of market dynamics with price expectations equivalent to a damped harmonic oscillator model, we analyze the reaction of financial markets to shocks. In order to do this, we gather data from indices of a variety of financial markets for the 1987 Black Monday, the Russian crisis of 1998, the crash after September 11th (2001), and the recent downturn of markets due to the subprime mortgage crisis in the USA (2008). Analyzing those data we were able to establish the amount by which each market felt the shocks, a dampening factor which expresses the capacity of a market of absorving a shock, and also a frequency related with volatility after the shock. The results gauge the efficiency of different markets in recovering from such shocks, and measure some level of dependance between them. \end{abstract}

\section{Introduction}

Market crashes are a phenomenon that has been gathering much attention in the past four years, particularly due to the ongoing subprime mortgage crisis, but it is a subject that has been attracting the attention of numerous researchers, particularly economists, for the last few decades.The scientific literature shows many attempts at modeling the transmission of volatility (contagion) between stock markets \cite{cont1}-\cite{cont8}, how the correlation between them change with time \cite{time1}-\cite{time3}, and how it tends to increase in times of crisis \cite{vol1}-\cite{vol13}. Recently, physicists have been building models trying to explain how and why such crashes occur. One particularly successful model (although not completely), the log-periodic model \cite{9801}-\cite{0801}, predicts crashes as the possible outcomes of increasingly fast oscillations due to the interaction between agents in a market.

In \cite{leocorr}, we have shown how stock markets tend to behave similarly in times of crisis, responding to a strong world market movement that may be the result of news shared by all markets or back reactions to other market's movements. This market movement increases in times of crisis and one may model individual stock markets' movements as the response of those markets to an external force (the market). 

Our work uses a modification of a simple model of market dynamics with price expectations which can be put into direct analogy with a model of a damped harmonic oscillator subject to some external force in order to study the behavior of some stock markets chosen to represent different parts of the world. Periods of crashes make their study simpler by the fact that often the large oscillations of the markets overshadow their typical random walk behavior. Linking the oscillations after a crash to a model of a damped harmonic oscillator also makes it possible to analyze some of the market's properties, such as resistance to change, correlation to the other markets, and volatility.

In order to do our work, we use data from stock market indices in 9 countries: USA, Hong Kong, Japan, Germany, UK, Brazil, Mexico, South Korea, and Australia, which, together, represent about 60\% of the world financial market capitalization (around 28 billion dollars per month). Two indices from the New York Stock Exchange (Dow Jones and S\&P 500) are used as benchmarks. The fits of data to the model are made using simple minimum squares techniques with an eye for the high nonlinearity of the function being calibrated.

\section{The choice of crises}

We now explain why we chose the crises of 1987, 1998, 2001, and 2008. Figure 1 shows the monthly average of the daily performances of the S\&P 500 index of the New York Stock Exchange from the beginning of 1980 to the middle of 2010. The crashes of 1987 (Black Monday), 1998 (Russian crisis), 2001 (September 11), and 2008 (subprime mortgage crisis) are outlined in order to show the periods we are focusing in our study of how markets react to strong shocks. These crises have been chosen because they were severe, global, and because they represent a variety of types. The crisis of 1987 was a surprise, probably caused by endogenous reasons; the 1998 Russian crisis was also abrupt, but followed by a quick recovery; the crisis of September 11 was an example of a crash that was caused almost completely by external news; and the crisis of 2008, which extended from 2007 up to 2009, was an example of a series of downfalls of the world markets.

\begin{pspicture}(-0.4,-0.7)(3.5,7)
\psset{xunit=0.038,yunit=0.004}
\psline{->}(0,0)(400,0) \psline{->}(0,0)(0,1600) \rput(415,0){year} \rput(25,1600){S\&P 500}\scriptsize \psline(0,-25)(0,25) \rput(0,-75){1980} \psline(61,-25)(61,25) \rput(61,-75){1985} \psline(122,-25)(122,25) \rput(122,-75){1990} \psline(183,-25)(183,25) \rput(183,-75){1995} \psline(244,-25)(244,25) \rput(244,-75){2000} \psline(304,-25)(304,25) \rput(304,-75){2005} \psline(364,-25)(364,25) \rput(364,-75){2010} \psline(-2.5,500)(2.5,500) \rput(-10,500){$500$} \psline(-2.5,1000)(2.5,1000) \rput(-12.5,1000){$1000$} \psline(-2.5,1500)(2.5,1500) \rput(-12.5,1500){$1500$}
\psline[linecolor=blue](0,114.16)(1,113.66)(2,102.09)(3,106.29)(4,111.24)(5,114.24)(6,121.67)(7,122.38)(8,125.46)
(9,127.47)(10,140.52)(11,135.76)(12,129.55)(13,131.27)(14,136)(15,132.81)(16,132.59)(17,131.21)(18,130.92)(19,122.79)
(20,116.18)(21,121.89)(22,126.35)(23,122.55)(24,120.4)(25,113.11)(26,111.96)(27,116.44)(28,111.88)(29,109.61)(30,107.09)
(31,119.51)(32,120.42)(33,133.72)(34,138.53)(35,140.64)(36,145.3)(37,148.06)(38,148.06)(39,152.96)(40,164.43)(41,162.39)
(42,167.64)(43,162.56)(44,164.4)(45,166.07)(46,163.55)(47,166.4)(48,164.93)(49,163.41)(50,157.06)(51,159.18)(52,160.05)
(53,150.55)(54,153.18)(55,150.66)(56,166.68)(57,166.1)(58,166.09)(59,163.58)(60,167.24)(61,179.63)(62,181.18)(63,180.66)
(64,179.83)(65,189.55)(66,191.85)(67,190.92)(68,188.63)(69,182.08)(70,189.82)(71,202.17)(72,211.28)(73,211.78)
(74,226.92)(75,238.9)(76,235.52)(77,247.35)(78,250.84)(79,236.12)(80,252.93)(81,231.32)(82,243.98)(83,249.22)
(84,242.17)(85,274.08)(86,284.2)(87,291.7)(88,288.36)(89,290.1)(90,304)(91,318.66)(92,329.8)(93,321.83)(94,251.79)
(95,230.3)(96,247.08)(97,257.07)(98,267.82)(99,258.89)(100,261.33)(101,262.16)(102,273.5)(103,272.02)(104,261.52)
(105,261.52)(106,271.91)(107,278.97)(108,273.7)(109,277.72)(110,297.47)(111,288.86)(112,294.87)(113,309.64)(114,320.52)
(115,317.98)(116,346.08)(117,351.45)(118,349.15)(119,340.36)(120,345.99)(121,353.4)(122,329.08)(123,331.89)(124,339.94)
(125,330.8)(126,361.23)(127,358.02)(128,356.15)(129,322.56)(130,306.05)(131,304)(132,322.22)(133,330.22)(134,343.93)
(135,367.07)(136,375.22)(137,375.34)(138,389.83)(139,371.16)(140,387.81)(141,395.43)(142,387.86)(143,392.45)(144,375.22)
(145,417.09)(146,408.78)(147,412.7)(148,403.69)(149,414.95)(150,415.35)(151,408.14)(152,424.21)(153,414.03)(154,417.8)
(155,418.68)(156,431.35)(157,435.71)(158,438.78)(159,443.38)(160,451.67)(161,440.19)(162,450.19)(163,450.53)(164,448.13)
(165,463.56)(166,458.93)(167,467.83)(168,461.79)(169,466.45)(170,481.61)(171,467.14)(172,467.14)(173,445.77)(174,450.91)
(175,456.5)(176,444.27)(177,458.26)(178,475.49)(179,462.71)(180,472.35)(181,453.69)(182,459.27)(183,470.42)(184,487.39)
(185,500.71)(186,514.71)(187,533.4)(188,544.75)(189,562.06)(190,561.88)(191,584.41)(192,581.5)(193,605.37)(194,615.93)
(195,636.02)(196,640.43)(197,645.5)(198,654.17)(199,669.12)(200,670.63)(201,639.95)(202,651.99)(203,687.33)(204,705.27)
(205,757.02)(206,740.74)(207,786.16)(208,790.82)(209,757.12)(210,801.34)(211,848.28)(212,885.14)(213,954.31)(214,899.47)
(215,947.28)(216,914.62)(217,955.4)(218,970.43)(219,980.28)(220,1049.34)(221,1101.75)(222,1111.75)(223,1090.82)
(224,1133.84)(225,1120.67)(226,957.28)(227,1017.01)(228,1098.67)(229,1163.63)(230,1229.23)(231,1279.64)(232,1238.33)
(233,1286.37)(234,1335.18)(235,1301.84)(236,1372.71)(237,1328.72)(238,1320.41)(239,1320.41)(240,1282.71)(241,1362.93)
(242,1388.91)(243,1469.25)(244,1394.46)(245,1366.42)(246,1498.58)(247,1452.43)(248,1420.6)(249,1454.6)(250,1430.83)
(251,1517.68)(252,1436.51)(253,1429.4)(254,1314.95)(255,1320.28)(256,1366.01)(257,1239.94)(258,1160.33)(259,1249.46)
(260,1255.82)(261,1224.38)(262,1211.23)(263,1133.58)(264,1040.94)(265,1059.78)(266,1139.45)(267,1148.08)(268,1130.2)
(269,1106.73)(270,1147.39)(271,1076.92)(272,1067.14)(273,989.82)(274,911.62)(275,916.07)(276,815.28)(277,885.76)
(278,936.31)(279,879.82)(280,855.7)(281,841.15)(282,848.18)(283,916.92)(284,963.59)(285,974.5)(286,990.31)(287,1008.01)
(288,995.97)(289,1050.71)(290,1058.2)(291,1111.92)(292,1131.13)(293,1144.94)(294,1126.21)(295,1107.3)(296,1120.68)
(297,1140.84)(298,1101.72)(299,1104.24)(300,1114.58)(301,1130.2)(302,1173.82)(303,1211.92)(304,1181.27)(305,1203.6)
(306,1180.59)(307,1156.85)(308,1191.5)(309,1191.33)(310,1234.18)(311,1220.33)(312,1228.81)(313,1207.01)(314,1249.48)
(315,1248.29)(316,1280.08)(317,1280.66)(318,1294.87)(319,1310.61)(320,1270.09)(321,1270.2)(322,1276.66)(323,1303.82)
(324,1335.85)(325,1377.94)(326,1400.63)(327,1418.3)(328,1438.24)(329,1406.82)(330,1420.86)(331,1482.37)(332,1530.62)
(333,1503.35)(334,1455.27)(335,1473.99)(336,1526.75)(337,1549.38)(338,1481.14)(339,1468.36)(340,1378.55)(341,1330.63)
(342,1322.7)(343,1385.59)(344,1400.38)(345,1280)(346,1267.38)(347,1282.83)(348,1166.36)(349,968.75)(350,896.24)
(351,903.25)(352,825.88)(353,735.09)(354,797.87)(355,872.81)(356,919.14)(357,919.32)(358,987.48)(359,1020.62)
(360,1057.08)(361,1036.19)(362,1095.63)(363,1115.1)(364,1073.87)(365,1104.49)(366,1169.43)(367,1186.69)(368,1089.41)
(369,1030.71)(370,1101.6)(371,1120.46)
\parametricplot[linecolor=red,linewidth=2pt]{0}{360}{t cos 13.16 mul 94 add t sin 125 mul 251.79 add}
\parametricplot[linecolor=red,linewidth=2pt]{0}{360}{t cos 13.16 mul 227 add t sin 125 mul 1017.01 add}
\parametricplot[linecolor=red,linewidth=2pt]{0}{360}{t cos 13.16 mul 264 add t sin 125 mul 1040.94 add}
\parametricplot[linecolor=red,linewidth=2pt]{0}{360}{t cos 13.16 mul 349 add t sin 125 mul 968.75 add}
\rput(94,540){1987}
\rput(94,450){(Black Monday)}
\rput(227,800){1998}
\rput(235,700){(Russian crisis)}
\rput(275,1240){2001}
\rput(287,1340){(September 11)}
\rput(349,640){2008}
\rput(349,540){(subprime mortgage crisis)}
\end{pspicture}

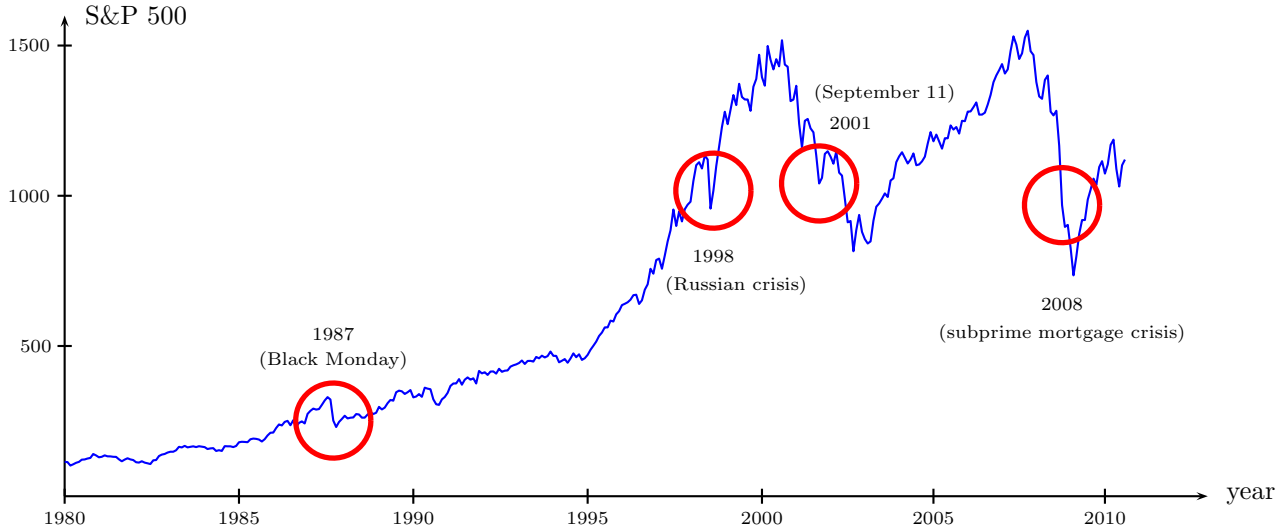
\noindent {\small Figure 1: S\&P 500 index (monthly averages), from the beginning of 1980 to the middle of 2010, with the chosen crises highlighted.}

\vskip 0.3 cm

It is well known that financial markets behave in good approximations of efficient markets, and most of the theory in finance is based on this assumption. That may be true for normal periods, but the efficient market hypothesis turns out to be wrong when markets are experiencing crises. Looking at the closing indices of every day in which there was negotiation, we consider now the log-returns of the S\&P 500 index, given by
\begin{equation}
S_t=\ln (P_t)-\ln (P_{t-1})\approx \frac{P_t-P_{t-1}}{P_t}\ .
\end{equation}

In order to best visualize the most extreme points, we next plot the log-frequency distribution (figure 2), defined as \begin{equation} \text{log-density}=\ln (1+\text{density})\ ,\end{equation}
where {\sl density} is the frequency distribution of the log-returns.

\begin{pspicture}(-10,-0.5)(3,4.6)
\psset{xunit=30,yunit=0.25} \psline{->}(-0.3,0)(0.2,0) \psline{->}(-0.3,0)(-0.3,16) \rput(0.23,0){interval} \rput(-0.26,16){log-density}
\scriptsize \psline(-0.3,-0.4)(-0.3,0.4) \rput(-0.3,-1.2){$-0.3$} \psline(-0.2,-0.4)(-0.2,0.4) \rput(-0.2,-1.2){$-0.2$} \psline(-0.1,-0.4)(-0.1,0.4) \rput(-0.1,-1.2){$-0.1$} \psline(0,-0.4)(0,0.4) \rput(0,-1.2){$0$} \psline(0.1,-0.4)(0.1,0.4) \rput(0.1,-1.2){$0.1$} \psline(-0.303,0)(-0.297,0) \rput(-0.315,0){0} \psline(-0.303,4)(-0.297,4) \rput(-0.315,4){$4$} \psline(-0.303,8)(-0.297,8) \rput(-0.315,8){$8$} \psline(-0.303,12)(-0.297,12) \rput(-0.315,12){$12$}
\pspolygon*[linecolor=lightred](-0.232,0)(-0.232,1.4395)(-0.228,1.4395)(-0.228,0)
\pspolygon[linecolor=red](-0.232,0)(-0.232,1.4395)(-0.228,1.4395)(-0.228,0)
\pspolygon*[linecolor=lightred](-0.096,0)(-0.096,2.8790)(-0.092,2.8790)(-0.092,0)
\pspolygon[linecolor=red](-0.096,0)(-0.096,2.8790)(-0.092,2.8790)(-0.092,0)
\pspolygon*[linecolor=lightred](-0.088,0)(-0.088,1.4395)(-0.084,1.4395)(-0.084,0)
\pspolygon[linecolor=red](-0.088,0)(-0.088,1.4395)(-0.084,1.4395)(-0.084,0)
\pspolygon*[linecolor=lightred](-0.08,0)(-0.08,1.4395)(-0.076,1.4395)(-0.076,0)
\pspolygon[linecolor=red](-0.08,0)(-0.08,1.4395)(-0.076,1.4395)(-0.076,0)
\pspolygon*[linecolor=lightred](-0.072,0)(-0.072,3.3425)(-0.068,3.3425)(-0.068,0)
\pspolygon[linecolor=red](-0.072,0)(-0.072,3.3425)(-0.068,3.3425)(-0.068,0)
\pspolygon*[linecolor=lightred](-0.064,0)(-0.064,3.3425)(-0.06,3.3425)(-0.06,0)
\pspolygon[linecolor=red](-0.064,0)(-0.064,3.3425)(-0.06,3.3425)(-0.06,0)
\pspolygon*[linecolor=lightred](-0.06,0)(-0.06,1.4395)(-0.056,1.4395)(-0.056,0)
\pspolygon[linecolor=red](-0.06,0)(-0.06,1.4395)(-0.056,1.4395)(-0.056,0)
\pspolygon*[linecolor=lightred](-0.056,0)(-0.056,2.8790)(-0.052,2.8790)(-0.052,0)
\pspolygon[linecolor=red](-0.056,0)(-0.056,2.8790)(-0.052,2.8790)(-0.052,0)
\pspolygon*[linecolor=lightred](-0.052,0)(-0.052,3.7211)(-0.048,3.7211)(-0.048,0)
\pspolygon[linecolor=red](-0.052,0)(-0.052,3.7211)(-0.048,3.7211)(-0.048,0)
\pspolygon*[linecolor=lightred](-0.048,0)(-0.048,2.2816)(-0.044,2.2816)(-0.044,0)
\pspolygon[linecolor=red](-0.048,0)(-0.048,2.2816)(-0.044,2.2816)(-0.044,0)
\pspolygon*[linecolor=lightred](-0.044,0)(-0.044,3.7211)(-0.04,3.7211)(-0.04,0)
\pspolygon[linecolor=red](-0.044,0)(-0.044,3.7211)(-0.04,3.7211)(-0.04,0)
\pspolygon*[linecolor=lightred](-0.04,0)(-0.04,4.5632)(-0.036,4.5632)(-0.036,0)
\pspolygon[linecolor=red](-0.04,0)(-0.04,4.5632)(-0.036,4.5632)(-0.036,0)
\pspolygon*[linecolor=lightred](-0.036,0)(-0.036,5.3268)(-0.032,5.3268)(-0.032,0)
\pspolygon[linecolor=red](-0.036,0)(-0.036,5.3268)(-0.032,5.3268)(-0.032,0)
\pspolygon*[linecolor=lightred](-0.032,0)(-0.032,7.1976)(-0.028,7.1976)(-0.028,0)
\pspolygon[linecolor=red](-0.032,0)(-0.032,7.1976)(-0.028,7.1976)(-0.028,0)
\pspolygon*[linecolor=lightred](-0.028,0)(-0.028,8.1244)(-0.024,8.1244)(-0.024,0)
\pspolygon[linecolor=red](-0.028,0)(-0.028,8.1244)(-0.024,8.1244)(-0.024,0)
\pspolygon*[linecolor=lightred](-0.024,0)(-0.024,9.2507)(-0.02,9.2507)(-0.02,0)
\pspolygon[linecolor=red](-0.024,0)(-0.024,9.2507)(-0.02,9.2507)(-0.02,0)
\pspolygon*[linecolor=lightred](-0.02,0)(-0.02,10.6537)(-0.016,10.6537)(-0.016,0)
\pspolygon[linecolor=red](-0.02,0)(-0.02,10.6537)(-0.016,10.6537)(-0.016,0)
\pspolygon*[linecolor=lightred](-0.016,0)(-0.016,11.7317)(-0.012,11.7317)(-0.012,0)
\pspolygon[linecolor=red](-0.016,0)(-0.016,11.7317)(-0.012,11.7317)(-0.012,0)
\pspolygon*[linecolor=lightred](-0.012,0)(-0.012,12.8939)(-0.008,12.8939)(-0.008,0)
\pspolygon[linecolor=red](-0.012,0)(-0.012,12.8939)(-0.008,12.8939)(-0.008,0)
\pspolygon*[linecolor=lightred](-0.008,0)(-0.008,14.0899)(-0.004,14.0899)(-0.004,0)
\pspolygon[linecolor=red](-0.008,0)(-0.008,14.0899)(-0.004,14.0899)(-0.004,0)
\pspolygon*[linecolor=lightred](-0.004,0)(-0.004,15.0683)(0,15.0683)(0,0)
\pspolygon[linecolor=red](-0.004,0)(-0.004,15.0683)(0,15.0683)(0,0)
\pspolygon*[linecolor=lightred](0,0)(0,15.2318)(0.004,15.2318)(0.004,0)
\pspolygon[linecolor=red](0,0)(0,15.2318)(0.004,15.2318)(0.004,0)
\pspolygon*[linecolor=lightred](0.004,0)(0.004,14.3972)(0.008,14.3972)(0.008,0)
\pspolygon[linecolor=red](0.004,0)(0.004,14.3972)(0.008,14.3972)(0.008,0)
\pspolygon*[linecolor=lightred](0.008,0)(0.008,13.2851)(0.012,13.2851)(0.012,0)
\pspolygon[linecolor=red](0.008,0)(0.008,13.2851)(0.012,13.2851)(0.012,0)
\pspolygon*[linecolor=lightred](0.012,0)(0.012,11.9925)(0.016,11.9925)(0.016,0)
\pspolygon[linecolor=red](0.012,0)(0.012,11.9925)(0.016,11.9925)(0.016,0)
\pspolygon*[linecolor=lightred](0.016,0)(0.016,10.7380)(0.02,10.7380)(0.02,0)
\pspolygon[linecolor=red](0.016,0)(0.016,10.7380)(0.02,10.7380)(0.02,0)
\pspolygon*[linecolor=lightred](0.02,0)(0.02,9.2984)(0.024,9.2984)(0.024,0)
\pspolygon[linecolor=red](0.02,0)(0.02,9.2984)(0.024,9.2984)(0.024,0)
\pspolygon*[linecolor=lightred](0.024,0)(0.024,7.9056)(0.028,7.9056)(0.028,0)
\pspolygon[linecolor=red](0.024,0)(0.024,7.9056)(0.028,7.9056)(0.028,0)
\pspolygon*[linecolor=lightred](0.028,0)(0.028,6.3228)(0.032,6.3228)(0.032,0)
\pspolygon[linecolor=red](0.028,0)(0.028,6.3228)(0.032,6.3228)(0.032,0)
\pspolygon*[linecolor=lightred](0.032,0)(0.032,6.2215)(0.036,6.2215)(0.036,0)
\pspolygon[linecolor=red](0.032,0)(0.032,6.2215)(0.036,6.2215)(0.036,0)
\pspolygon*[linecolor=lightred](0.036,0)(0.036,5.7581)(0.04,5.7581)(0.04,0)
\pspolygon[linecolor=red](0.036,0)(0.036,5.7581)(0.04,5.7581)(0.04,0)
\pspolygon*[linecolor=lightred](0.04,0)(0.04,4.0412)(0.044,4.0412)(0.044,0)
\pspolygon[linecolor=red](0.04,0)(0.04,4.0412)(0.044,4.0412)(0.044,0)
\pspolygon*[linecolor=lightred](0.044,0)(0.044,3.3425)(0.048,3.3425)(0.048,0)
\pspolygon[linecolor=red](0.044,0)(0.044,3.3425)(0.048,3.3425)(0.048,0)
\pspolygon*[linecolor=lightred](0.048,0)(0.048,4.3186)(0.052,4.3186)(0.052,0)
\pspolygon[linecolor=red](0.048,0)(0.048,4.3186)(0.052,4.3186)(0.052,0)
\pspolygon*[linecolor=lightred](0.052,0)(0.052,2.2816)(0.056,2.2816)(0.056,0)
\pspolygon[linecolor=red](0.052,0)(0.052,2.2816)(0.056,2.2816)(0.056,0)
\pspolygon*[linecolor=lightred](0.06,0)(0.06,2.2816)(0.064,2.2816)(0.064,0)
\pspolygon[linecolor=red](0.06,0)(0.06,2.2816)(0.064,2.2816)(0.064,0)
\pspolygon*[linecolor=lightred](0.064,0)(0.064,1.4395)(0.068,1.4395)(0.068,0)
\pspolygon[linecolor=red](0.064,0)(0.064,1.4395)(0.068,1.4395)(0.068,0)
\pspolygon*[linecolor=lightred](0.084,0)(0.084,1.4395)(0.088,1.4395)(0.088,0)
\pspolygon[linecolor=red](0.084,0)(0.084,1.4395)(0.088,1.4395)(0.088,0)
\pspolygon*[linecolor=lightred](0.1,0)(0.1,1.4395)(0.104,1.4395)(0.104,0)
\pspolygon[linecolor=red](0.1,0)(0.1,1.4395)(0.104,1.4395)(0.104,0)
\pspolygon*[linecolor=lightred](0.108,0)(0.108,1.4395)(0.112,1.4395)(0.112,0)
\pspolygon[linecolor=red](0.108,0)(0.108,1.4395)(0.112,1.4395)(0.112,0)
\parametricplot[linecolor=blue,linewidth=2pt]{0}{360}{t cos 0.02 mul -0.230 add t sin 2 mul 1.4395 add}
\parametricplot[linecolor=blue,linewidth=2pt]{0}{360}{t cos 0.02 mul -0.078 add t sin 2 mul 1.4395 add}
\parametricplot[linecolor=blue,linewidth=2pt]{0}{360}{t cos 0.02 mul -0.050 add t sin 2 mul 3.7211 add}
\parametricplot[linecolor=blue,linewidth=2pt]{0}{360}{t cos 0.02 mul -0.094 add t sin 2 mul 2.8790 add}
\end{pspicture}

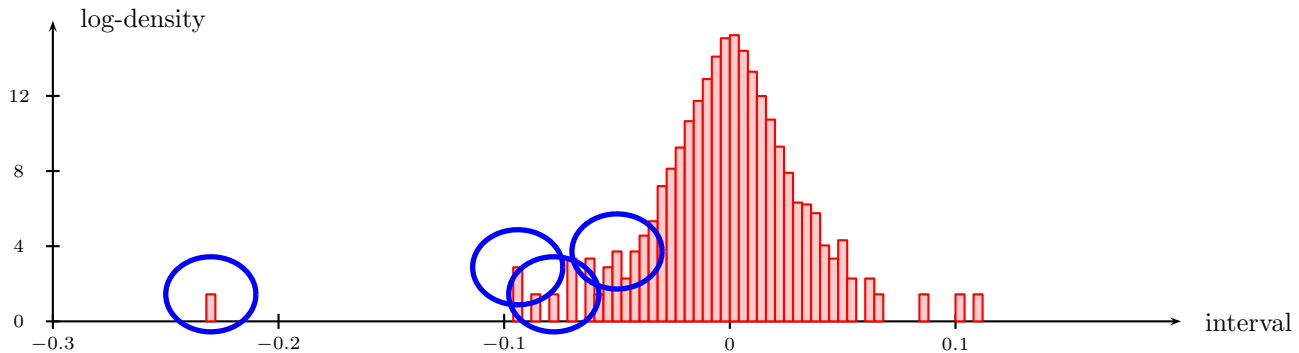
\hskip 1 cm {\small Figure 2: log-density of log-returns of the S\&P 500 index, computed from the beginning of 1980 to the end of 2008.}

\vskip 0.3 cm

If the probability distribution function relative to the log-returns were a Gaussian, then we were not not expect large deviations from the average of the log-returns. Nevertheless, one can clearly see from figure 2 that there are strong deviations, mainly towards negative log-returns, that cannot be explained by a Gaussian distribution. In figure 2, a Gaussian distribution function should be represented by a parabola, what is clearly not the case, so financial crises should not be studied using the efficient market hypothesis. In figure 2, we circled the densitites relative to the four crises we are studying, showing that they correspond to diverse degrees of deviation from the norm.

\section{The model}

Now we shall explain the kind of dynamic equations we are using in our model, and why, starting by connecting them with the equations of a well known model of economic dynamics.

\subsection{Market dynamics with price expectation}

We shall begin with a very simple model of price change due to market demands, which is used in textbooks on economic dynamics \cite{ED1}. The model is based on the following two equations of demanded quantity $Q_d$ and supplied quantity $Q_s$ as functions of the price $P$ of a commodity and its first and second order derivatives:
\begin{eqnarray}
\label{dem}
 & & Q_d=\alpha_0-\alpha_1P+\alpha_2\dot P-\alpha_3\ddot P\ ,\\
\label{sup}
 & & Q_s=\beta_0+\beta_1P-\beta_2\dot P+\beta_2\ddot P\ ,
\end{eqnarray}
where all constants are considered real and positive, and where the dynamics is given by
\begin{equation}
\label{dyn}
\dot P=\lambda (Q_d-Q_s)\ ,
\end{equation}
where $\lambda >0$. Here $\dot P$ stands for the derivative of $P$ with respect to time $t$.

Equation (\ref{dem}) expresses the assumption that demand falls when price rises, and that demand grows when the market perceives that prices have an upward tendency, and falls when this tendency is decreasing. In a similar way, equation (\ref{sup}) expresses that supply rises when price goes up, falls when there is a further tendency of the price rising, and goes up when this tendency is growing. The dynamic equation (\ref{dyn}) says that, when demand is greater than supply, prices tend to rise, and when supply is higher than demand, prices tend to drop.

Such a model is supported by some experimental evidence \cite{B1} and is considered appropriate for market dynamics, supposing there is not too much noise. It leads to the following differential equation for the price as a function of time:
\begin{equation}
\label{eqprice}
\lambda (\alpha_1+\beta_1)\ddot P+\left[ 1-\lambda (\alpha_2+\beta_2)\right] \dot P+\lambda (\alpha_1+\beta_1)P=\lambda (\alpha_0-\beta_0)\ .
\end{equation}

This differential equation has three types of solutions, depending on the parameters of the model, specifically, on the combination
\begin{equation}
\label{delta}
\Delta =4\lambda^2(\alpha_3+\beta_3)(\alpha_1+\beta_1)-\left[ 1-\lambda (\alpha_2+\beta_2)\right] ^2 \ .
\end{equation}

If $\Delta >0$, then
\begin{equation}
\label{solprice1}
P(t)=c_1\e^{r_1t}+c_2\e^{r_2t}+P^*\ ,
\end{equation}
being $c_1$ and $c_2$ constants,
\begin{equation}
r_1=\frac{-\left[ 1-\lambda (\alpha_2+\beta_2)\right] -\sqrt{\Delta }}{2\lambda (\alpha_3+\beta_3)}\ \ ,\ \ r_2=\frac{-\left[ 1-\lambda (\alpha_2+\beta_2)\right] +\sqrt{\Delta }}{2\lambda (\alpha_3+\beta_3)}\ ,
\end{equation}
and
\begin{equation}
\label{solpart}
P^*=\frac{\alpha_0-\beta_0}{\alpha_1+\beta_1}\ .
\end{equation}
This solution describes exponencial growth or exponential growth or fall up or down to an asymptotic limit.

The second solution occurs when $\Delta =0$:
\begin{equation}
\label{solprice2}
P(t)=c_1\e^{rt}+c_2t\e^{rt}+P^*\ ,
\end{equation}
where $c_1$ and $c_2$ are constants,
\begin{equation}
r=\frac{-\left[ 1-\lambda (\alpha_2+\beta_2)\right] }{2\lambda (\alpha_3+\beta_3)}\ ,
\end{equation}
and $P^*$ is given by (\ref{solpart}). This solution broadly describes the same situation as the first type of solution.

The third solution occurs when $\Delta <0$, and is given by
\begin{equation}
\label{solprice3}
P(t)=c_1\e ^{-\beta t}\cos (wt-\varphi )+P^*\ ,
\end{equation}
where $c_1$ (the amplitude) and $\varphi $ (the phase) are constants,
\begin{eqnarray}
\label{coef1}
 & & \beta =\frac{1-\lambda (\alpha_2+\beta _2)}{2\lambda (\alpha_3+\beta_3)}\ ,\\
\label{coef2}
 & & w=\frac{\sqrt{-\Delta }}{2(\alpha_3+\beta_3)}=\frac{\sqrt{\left[ 1-\lambda (\alpha_2+\beta_2)\right] ^2-4\lambda^2(\alpha_3+\beta_3)(\alpha_1+\beta_1)}}{2(\alpha_3+\beta_3)}\ ,
\end{eqnarray}
and $P^*$ is given by (\ref{solpart}). Depending on $\beta $, this equation describes an oscillatory movement that diminishes exponentially with time or increases exponentially in time.

The following three figures (figures 3a, 3b, and 3c) illustrate the three types of solutions. As can be seen, $P^*$ is the price of equilibrium for the three solutions, where $P_0$ is the starting price.

\begin{pspicture}(-0.5,-0.5)(3,4.5)
\psset{xunit=1.3,yunit=1.3} \psline{->}(0,0)(3,0) \psline{->}(0,0)(0,3) \small
\rput(0.2,3){$P$} \rput(3.2,0){$t$} \psline(-0.1,1)(0.1,1) \psline(-0.1,2)(0.1,2) \small \rput(-0.2,-0.2){0} \rput(-0.4,1){$P^*$} \rput(-0.4,2){$P_0$} \psplot[linecolor=red]{0}{2.8}{2.71828182 x -4 mul exp -1.5 mul 2.71828182 x -2 mul exp 2.5 mul add 1 add}
\end{pspicture}
\begin{pspicture}(-3,-0.5)(3,4.5)
\psset{xunit=1.3,yunit=1.3} \psline{->}(0,0)(3,0) \psline{->}(0,0)(0,3) \small
\rput(0.2,3){$P$} \rput(3.2,0){$t$} \psline(-0.1,1)(0.1,1) \psline(-0.1,2)(0.1,2) \small \rput(-0.2,-0.2){0} \rput(-0.4,1){$P^*$} \rput(-0.4,2){$P_0$} \psplot[linecolor=red]{0}{2.8}{2.71828182 x -4 mul exp 1 mul 2.71828182 x -4 mul exp x mul 4 mul add 1 add}
\end{pspicture}
\begin{pspicture}(-3,-0.5)(3,4.5)
\psset{xunit=1.3,yunit=1.3} \psline{->}(0,0)(3,0) \psline{->}(0,0)(0,3) \small
\rput(0.2,3){$P$} \rput(3.2,0){$t$} \psline(-0.1,1)(0.1,1) \psline(-0.1,2)(0.1,2) \small \rput(-0.2,-0.2){0} \rput(-0.4,1){$P^*$} \rput(-0.4,2){$P_0$} \psplot[linecolor=red]{0}{2.8}{2.71828182 x -2 mul exp x 400 mul cos 1 mul mul 1 add}
\end{pspicture}

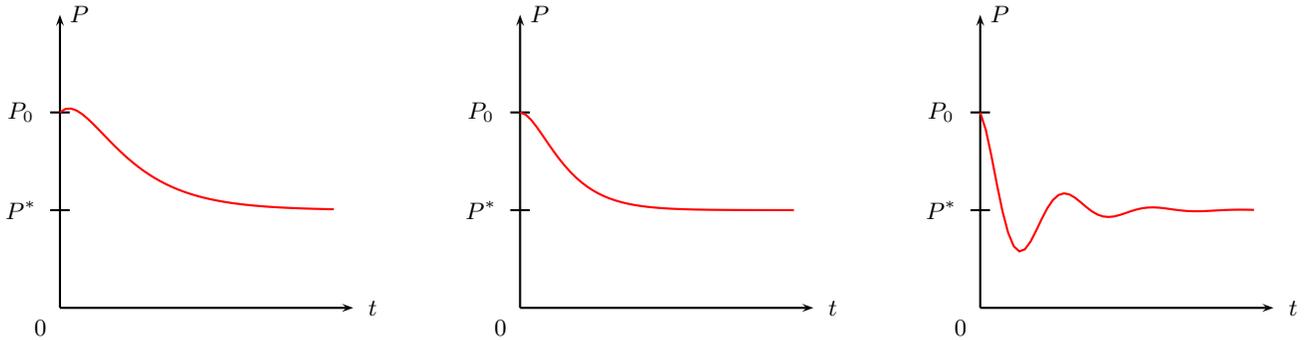
\centerline{\small \hskip 0.2 cm Figure 3: a) example of solution for $\Delta >0$. \hskip 0.2 cm b) example of solution for $\Delta =0$. \hskip 1 cm c) example of solution for $\Delta <0$.}

\vskip 0.3 cm

So now one may ask why such a simple model should be introduced here, and what does it have to do with financial market crashes. If one looks at figures 3a and 3b, one can see that it resembles the behavior of market indices when there is a slow drop in prices, given or taken some noise. Now, the third figure ilustrates aproximately what happens to markets when there is a sharp drop (a crash), followed by oscillations due to speculation which further disappear into the usual noise close to a price $P^*$ that the market considers more appropriate.

Our insterest in this article is to analyze the third type of behavior, and to gauge how each financial market behaved in previous crises when viewed through the light of a model similar to this, although the model we shall be using is not exactly this one, as shall be explained now.

When looking at a differential equation like (\ref{eqprice}), one must consider the homogeneous part of this equation (left side of that particular one) to describe endogenous effects, due to the nature of the markets. The nonhomogeneus part (right side of that equation) describes exogenous effects. What we shall do now is add another term, $\delta \e^{-\alpha t}$, which describes an exogenous shock in the market that is very sharp in the begining, and that drops very rapidly with time. Such a shock tries to mimic the effect of some external information that may well be information about the drops of other financial markets or the effect of endogenous phase transitions, in a particular market. Doing so for differential equation (\ref{eqprice}), one obtains

\begin{equation}
\label{eqpricemod}
\lambda (\alpha_1+\beta_1)\ddot P+\left[ 1-\lambda (\alpha_2+\beta_2)\right] \dot P+\lambda (\alpha_1+\beta_1)P=\lambda (\alpha_0-\beta_0)+\delta \e^{-\alpha t}\ ,
\end{equation}
where all constants are considered to be real and positive.

The solution for this differential equation when there is an oscillatory behavior is
\begin{equation}
\label{solprice}
P(t)=c_1\e ^{-\beta t}\cos (wt-\varphi )+P^*+b\e^{-\alpha t}\ ,
\end{equation}
where
\begin{equation}
b=\frac{\delta }{\lambda (\alpha_3+\beta_3)\alpha^2-\left[ 1-\lambda (\alpha_2+\beta_2)\right] \alpha +\alpha_1+\beta_1}\ ,
\end{equation}
and the other coefficients are given by equations (\ref{coef1}) and (\ref{coef2}).

Before using this model in order to gauge real financial data, it is best if one first sees it through the eyes of another theory that originally comes from physics.

\subsection{Damped harmonic oscillator}

The model we have just described can be put in direct analogy with another model of a damped harmonic oscillator that makes it easier to comprehend the solutions we have just shown in the last subsection.

We start by considering a body suspended by a spring. This body is subject to two forces: the weight $\vec P$ and the traction $\vec T$ exerted by the spring. When both forces are evenly matched, the body stays at the equilibrium point $x=0$. When an external force is applied to it, then the body will be displaced by a length $x$, and a restoring force $\vec F$ will appear.

This type of force is described by Hook's law,
\begin{equation}
\label{hook}
F=-kx\Leftrightarrow m\ddot x=-kx\ ,
\end{equation}
where $a=\ddot x$ is the acceleration, $m$ is the body's mass, and $k$ is the spring's constant, which depends on how the spring deforms. The solution to this differencial equation is a combination of sines and cosines, which means it is a purely oscillatory solution.

If one adds the term $-\gamma \dot x$ to the right side of equation (\ref{hook}), then one obtains
\begin{equation}
\label{damped}
m\ddot x=-kx-\gamma \dot x\ ,
\end{equation}
which is the equation of a damped harmonic oscillator. The solutions to this equation range from exponencial decay to oscillations that decay exponentially, just like the solutions for the market model with price expectations.

We now consider a model given by
\begin{equation}
\label{oscil1}
m\ddot P=-k(P-P^*)-\gamma \dot P+\delta \e^{-\alpha t}\ ,
\end{equation}
where $P$ is the price of a stock, $P-P^*$ is the distance between the price and a price $P^*$ that the market considers to be fair, and $\delta \e^{-\alpha t}$ is the shock we have introduced in the last subsection. This equation may be written like
\begin{equation}
\label{oscil}
m\ddot P+\gamma \dot P+kP=kP^*+\delta \e^{-\alpha t}\ ,
\end{equation}
which is essentially the same equation as (\ref{eqpricemod}), and has the following possible solutions:
\begin{eqnarray}
 & & P(t)=c_1\e^{-r_1t}+c_2\e^{-r_2t}+P^*+b\e^{-\alpha t},\ r_1=\frac{\gamma +\sqrt{\gamma ^2-4mk}}{2m},\ r_2=\frac{\gamma -\sqrt{\gamma ^2-4mk}}{2m},\ \Delta >0\ ;\\
 & & P(t)=c_1\e^{-rt}+c_2t\e^{-rt}+P^*+b\e^{-\alpha t}\ \ ,\ \ r=\frac{\gamma }{2m}\ \ ,\ \ \Delta =0\ ;\\
\label{soldamped}
 & & P(t)=c_1\e^{-\beta t}\cos (wt-\varphi )+P^*+b\e^{-\alpha t}\ \ ,\ \ \beta =\frac{\gamma }{2m}\ ,\ w=\frac{\sqrt{4mk-\gamma ^2}}{2m}\ \ ,\ \ \Delta <0\ ,
\end{eqnarray}
where $c_1$, $c_2$, and $\varphi $ are constants and
\begin{equation}
\label{deltab}
\Delta =\gamma ^2-4mk\ \ ,\ \ b=\frac{\delta }{m\alpha ^2-\gamma \alpha +k}\ .
\end{equation}

Solution (\ref{soldamped}) is particularly interesting to us. It describes a damped harmonic oscillation of the body attached to the spring when subject to an external force that is very strong at the first moment, and then diminishes quickly. The solution may be compared directly with solution (\ref{solprice}) if one sets
\begin{equation}
m=\lambda (\alpha_3+\beta_3)\ \ ,\ \ \gamma =1-\lambda (\alpha_2+\beta_2)\ \ ,\ \ k=\lambda (\alpha_1+\beta_1)\ .
\end{equation}

Thus, one can now make analogies and have a more physical picture of the model of market dynamics. Coefficients $\alpha_3$ and $\beta_3$, which are the influence the accelleration of prices have on the variation of price, may be seen as the ``mass'' of the system (seasoned by constant $\lambda $). Mass has the effect of inertia: a system with high mass is more resistant to shocks, but is also slower to settle after it has been put into motion. Coefficients $\alpha_2$ and $\beta_2$, which gauge the dependance of $\dot P$ on itself (when multiplied by $\lambda $), are the equivalent to the dampening factor $\gamma $. The coefficients $\alpha_1$ and $\beta_1$, which are related with the dependance of $\dot P$ with the price $P$, are essentially the constant of the spring, and determine how strong the spring is.

In what follows, we shall consider the oscillatory solution to be given by
\begin{equation}
\label{sol}
P(t)=A+B\e^{-\alpha t}+C\e^{-\beta t}\cos (wt-\varphi )\ ,
\end{equation}
where $A$, $B$, $\alpha $, $C$, $\beta $, $w$ and $\varphi $ are all parameters determined by the experimental data. Constant $A$ establishes the average axis around which the solution oscillates, and is also its assymptotic solution; constants $B$ and $\alpha $ determine the strength of the initial shock and how fast it decreases; constant $C$ determines the amplitudes of the oscillations; constant $\beta $, how fast those oscillations diminish with time; constant $w$, the frequency, how fast it oscillates (related with volatility after a crash); and $\varphi $ is the phase, which determines where are the peaks of the oscillations. The price $P$ will be exchanged for indices of markets, like the Dow Jones (USA) or the Hang Seng (Hong Kong).

This model has no less than seven parameters, which must be determined by real data on market indices, and most of them are linked in a nonlinear way. Section 5 explains how we attempted to do that.

\section{Relation with the log-periodic model}

Sornette and Johansen, followed by other collaborators, developed a model for the behavior of financial markets prior to and after crashes \cite{9801}-\cite{0801}, as well as other economic systems \cite{other01}, \cite{other02}. This model has been extensively tested among the years (for recent, yet unpublished articles, see \cite{new01}-\cite{new04}) on a great number of markets, and has also received a solid theoretical standing based on the theory of phase transitions and finance theory, also receiving a certain number of criticism \cite{crit01}, \cite{crit02}. Their model, in its simplest form, can be represented by the following equations:
\begin{equation} \label{logprior} P(t)=A+B(t_c-t)^\beta +C(t_c-t)^\beta \cos \left( w\ln (t_c-t)-\phi \right) \end{equation}
prior to a crisis and
\begin{equation} \label{logafter} P(t)=A+B(t-t_c)^\beta +C(t-t_c)^\beta \cos \left( w\ln (t-t_c)-\phi \right) \end{equation}
after a crash. In these equations, $t_c$ is the critical time at which the crash occurs, where the model goes to infinity and loses validity. Their model is measured in months and years, and the oscillations get stronger as time approaches the critical value.

Using the change of variable
\begin{equation}
\tau =\ln (t_c-t)\ ,
\end{equation}
into equation (\ref{logafter}), one has
\begin{equation} P(t)=A+B\e ^{\beta \tau }+C\e ^{\beta \tau }\cos \left( w\tau -\phi \right) \ ,\end{equation}
which can be readily compared with equation (\ref{sol}), with the important difference that we have two distinct values for the coefficients of the two exponencials. Considering now that their model measures time in terms of months or years and ours measures time in terms of a few days, one can use the expansion $\ln x\approx (x-k)+(1/2)(x-k)^2$ around a $x_0=k$ to establish that
\begin{equation}
\tau \approx t-t_c\ ,
\end{equation}
so that our model can be seen as a kind of approximation of the log-periodic model for times close to $t_c$.

This is an important aspect of our model, and we shall go back to this comparison after we start analyzing our results coming from real data.

\section{Data}

Let's now explain how the data were treated. First of all, in order to compare the various indices, we worked with normalized values of them, given by the indices divided by their average along the period being studied. This was not a problem in our case, for we only considered very small time intervals.

We next show how the fit to equation (\ref{sol}) was made. Since we have seven parameters to ajust, and the function to be ajusted is highly nonlinear, the straight use of the minimum squares method usually leads to disaster. The use of numercial methods is also not very reliable, since the landscape of the error function is rather complex and those algorithms tend to look for the closest minimum. We are going to use the index for the Bovespa (São Paulo stock exchange, in Brazil) during the crash of 1987 in order to illustrate the procedure we used, for the fit is relatively complex compared with some others.

First of all, we tried to remove the dependance on $A$, what just corresponds to reescaling the data. Using least squares method, which corresponds to minimizing the error function
\begin{equation}
E=\sum_{i=1}^n\left[ P(i)-P_i\right] ^2\ ,
\end{equation}


\vskip 0.1 cm

\noindent \begin{minipage}{10.4 cm }
where $P(i)$ is given by (\ref{sol}) and $P_i$ are the real data, we first find $A=1$ (as expected, since the data are normalized). Second, we removed a term $B\exp(-\alpha t)$, and minimized the difference of the real data with this term subtracted from it, resulting in $B=9.34$ and $\alpha =0.36$. Third, we removed a term $C\exp(-\beta t)$ and the resulting term $\cos (wt-\varphi )$, obtaining $C=0.14$, $\beta =0.22$, $w=2.92$, and $\varphi =1.52$.

\hskip 0.5 cm From this starting point, we tried to adjust the parameters again independently, using least squares method, and then all the parameters together, so that the theoretical curve resembles as best as we can the real data. The parameters for this curve are $A=0.92$, $B=8.1$, $\alpha =0.67$, $C=0.21$, $\beta=0.15$, $w=1.14$, and $\varphi =-9.1$.
\end{minipage}
\begin{pspicture}(-1.5,7)(1,7)
\psset{xunit=0.2,yunit=7} \psaxes[subticks=5,Dx=5,Dy=0.1,Ox=0,Oy=1.4]{->}(0,0.8)(28,1.4) \rput(29,0.8){$t$} \rput(1,1.4){$v$}
\psdots[linecolor=blue](1,1.25)(2,1.22)(3,1.27)(4,1.28)(5,1.21)(6,1.01)(7,1.03)(8,0.95)(9,0.95)(10,0.86)(11,0.88)
(12,0.89)(13,0.94)(14,0.94)(15,0.93)(16,0.91)(17,0.91)(18,0.95)(19,0.95)(20,0.99)(21,0.98)(22,0.99)(23,0.99)(24,0.95)
(25,0.94)(26,0.92)(27,0.91)
\psline(1,1.25)(2,1.22)(3,1.27)(4,1.28)(5,1.21)(6,1.01)(7,1.03)(8,0.95)(9,0.95)(10,0.86)(11,0.88)
(12,0.89)(13,0.94)(14,0.94)(15,0.93)(16,0.91)(17,0.91)(18,0.95)(19,0.95)(20,0.99)(21,0.98)(22,0.99)(23,0.99)(24,0.95)
(25,0.94)(26,0.92)(27,0.91)
\psplot[linecolor=red,plotpoints=500]{4.2}{28}{2.72828182 x -0.67 mul exp 8.1 mul 0.92 add 1.14 x mul -9.1 add 57.2958 mul cos 0.21 mul 2.72828182 x -0.15 mul exp mul add }
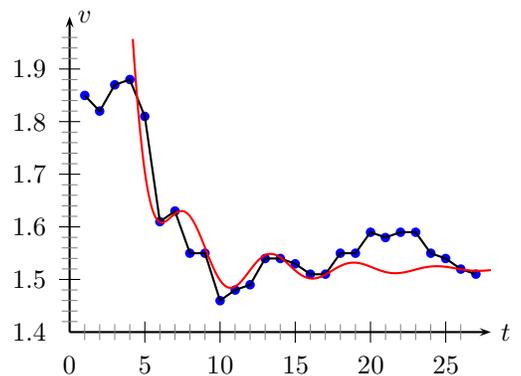
\rput(13,0.65){\small Figure 4: first fit of the Ibovespa for the crash of 1987.}
\end{pspicture}

\vskip 0.5 cm

After that, the error function can be plotted in terms of each of the many parameters while keeping the remaining parameters constant, in the vicinity of the parameters we obtained. There are clear values for the best fits of parameters $A$, $B$, $\alpha $, $C$, and $\beta $, in the vicinity of the values that have been chosen. For the phase $\varphi$, the best fit oscillates, as was to be expected. Now, for the parameter $w$, the function is more complex, and we can see minima near 0.6 and 2.8, being 2.8 a lower minimum. We shall not consider higher values for $w$, because for large enough values of the frequency, all parameters can be adjusted at will, but without economic meaning.

After we fix the value of $w$, we may now do some fine tunning minimizing the error function for the remaining parameters. This implies in changes in the value of $w$, which in turn leads to changes in the remaining parameters. This process can go on for a while, and it normally tends to converge, as long as the variations of $w$ are done separately. The two graphs bellow show the fits obtained when one leaves from the values $w=0.6$ and $w=2.8$. Leaving from $w=0.6$, the best fit is $A=0.9373$, $B=6.0237$, $\alpha =0.6976$, $C=0.0541$, $\beta =0.0414$, $w=0.8016$, and $\varphi=-1.3423$. Leaving from $w=2.8$, the best fit is $A=0.9355$, $B=6.2195$, $\alpha =0.6788$, $C=0.1948$, $\beta =0.2308$, $w=2.8555$, and $\varphi=1.2748$.

\vskip 0.2 cm

\begin{pspicture}(-1.5,5.4)(5.8,10)
\psset{xunit=0.2,yunit=7} \psaxes[subticks=5,Dx=5,Dy=0.1,Ox=0,Oy=1.4]{->}(0,0.8)(28,1.4) \rput(29,0.8){$t$} \rput(1,1.4){$v$}
\psdots[linecolor=blue](1,1.25)(2,1.22)(3,1.27)(4,1.28)(5,1.21)(6,1.01)(7,1.03)(8,0.95)(9,0.95)(10,0.86)(11,0.88)
(12,0.89)(13,0.94)(14,0.94)(15,0.93)(16,0.91)(17,0.91)(18,0.95)(19,0.95)(20,0.99)(21,0.98)(22,0.99)(23,0.99)(24,0.95)
(25,0.94)(26,0.92)(27,0.91)
\psline(1,1.25)(2,1.22)(3,1.27)(4,1.28)(5,1.21)(6,1.01)(7,1.03)(8,0.95)(9,0.95)(10,0.86)(11,0.88)
(12,0.89)(13,0.94)(14,0.94)(15,0.93)(16,0.91)(17,0.91)(18,0.95)(19,0.95)(20,0.99)(21,0.98)(22,0.99)(23,0.99)(24,0.95)
(25,0.94)(26,0.92)(27,0.91)
\psplot[linecolor=red,plotpoints=500]{3.7}{28}{2.72828182 x -0.6976 mul exp 6.0237 mul 0.9373 add 0.8016 x mul 1.3423 add 57.2958 mul cos 0.0541 mul 2.72828182 x -0.0414 mul exp mul add }
\rput(11,0.65){\small Figure 5: a) best fit for $w=0.8016$.}
\end{pspicture}
\begin{pspicture}(-2.5,5.4)(5.8,10)
\psset{xunit=0.2,yunit=7} \psaxes[subticks=5,Dx=5,Dy=0.1,Ox=0,Oy=1.4]{->}(0,0.8)(28,1.4) \rput(29,0.8){$t$} \rput(1,1.4){$v$}
\psdots[linecolor=blue](1,1.25)(2,1.22)(3,1.27)(4,1.28)(5,1.21)(6,1.01)(7,1.03)(8,0.95)(9,0.95)(10,0.86)(11,0.88)
(12,0.89)(13,0.94)(14,0.94)(15,0.93)(16,0.91)(17,0.91)(18,0.95)(19,0.95)(20,0.99)(21,0.98)(22,0.99)(23,0.99)(24,0.95)
(25,0.94)(26,0.92)(27,0.91)
\psline(1,1.25)(2,1.22)(3,1.27)(4,1.28)(5,1.21)(6,1.01)(7,1.03)(8,0.95)(9,0.95)(10,0.86)(11,0.88)
(12,0.89)(13,0.94)(14,0.94)(15,0.93)(16,0.91)(17,0.91)(18,0.95)(19,0.95)(20,0.99)(21,0.98)(22,0.99)(23,0.99)(24,0.95)
(25,0.94)(26,0.92)(27,0.91)
\psplot[linecolor=red,plotpoints=500]{3.65}{28}{2.72828182 x -0.6788 mul exp 6.2195 mul 0.9355 add 2.8555 x mul -1.2748 add 57.2958 mul cos 0.1948 mul 2.72828182 x -0.2308 mul exp mul add }
\rput(10,0.65){\small b) best fit for $w=2.8555$.}
\end{pspicture}

\vskip 1.2 cm

What one can see is that, for $w=0.8016$, the best fit ignores the first oscillation and is better adjusted to the last ones. For $w=2.8555$, the curve fits well the first oscillations, becoming increasingly innacurate for the last ones.

Before going into the detailed analysis of each index, we must explain which indices we chose, and why we have done so. First, we chose two distinct indices of the New York Exchange, the Dow Jones and the S\&P 500, in order to compare the same financial market when viewed by two different angles. It also serves for us to gauge our ability to correctly compute the parameters of the model. We also chose the Nasdaq, which reflects a more technologically oriented market. The Hang Seng of the Hong Kong Exchange, and the Nikkei, of the Tokyo Stock Exchange, were chosen in order to represent the Asian markets. We chose DAX (Germany) and FTSE (UK) as representatives of the European market, and Ibovespa (Brazil), and IPC (Mexico) as representatives of Latin America. Finally, we chose Kospi (South Korea) and ASX (Australia) as representing other emerging markets.

\subsection{The crash of 1987 - Black Monday}

Monday, October 19, 1987, was a day of shocked investors, baffled analysts, and despaired traders. The collapse started in Hong Kong, and swept its way towards the west. By the end of the month, markets throughout the world had fallen more than 30\%, in average. The world took many years to recover.

It is still not well known what were the reasons for the crash of October, 1987. Some point to a failure in the computer operated systems, some to panic generated by the fear of another 1929, some to external economic influences. What is known is that the efficient market hypothesis failed terribly in that day, and in the days that followed.

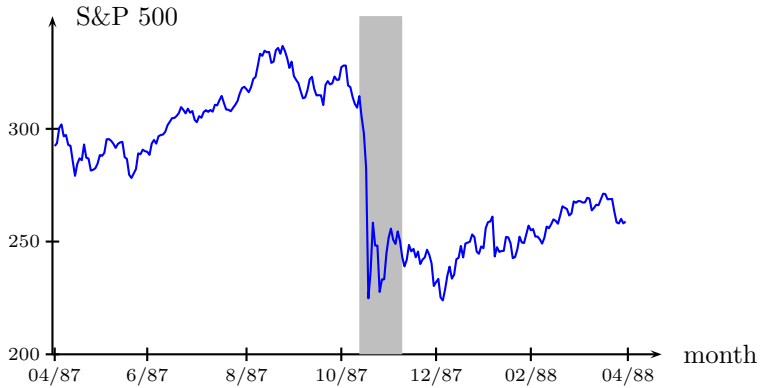
Figure 6 shows the Dow Jones index from April/1987 to April/1988. The period from 10/13/1987 to 11/08/1987 is highlighted, showing the time span we are considering for our model.



\vskip 0.7 cm

Figure 7 shows the graphs of the best fits that were obtained for the 1987 crash. The error function was calculated for the first 12 days (from 10/13/1987 to 11/08/1987), thus giving a maximum priority to the fitting of the first few oscillations after the crash. This is an example of an endogenous crash without a quick recovery: there were no news that triggered the downfall of all financial markets in the world, but just a sequence of small bad news, and the news of the markets themselves.

\vskip 1.8 cm

%

\vskip 3.4 cm

The parameters are displayed together in the next table.
\[ \begin{array}{|c||c|c|c|c|c|c|c|c|c|c|c|} \hline \text{Parameter} & \text{Dow Jones} & \text{S\&P 500} & \text{Nasdaq} & \text{Hang Seng} & \text{Nikkei} & \text{Dax} & \text{FTSE} & \text{Ibovespa} & \text{IPC} & \text{Kospi} & \text{Asx} \\ \hline A & 0.97 & 0.97 & 0.94 & 0.87 & 0.97 & 0.83 & 0.91 & 0.94 & 0.35 & -0.02 & 0.90 \\ B & 0.59 & 0.82 & 0.76 & 4.09 & 1.14 & 0.58 & 0.95 & 7.01 & 1.80 & 1.07 & 1.90 \\ \alpha & 0.70 & 0.68 & 0.36 & 0.51 & 0.53 & 0.11 & 0.29 & 0.70 & 0.08 & 0.00 & 0.40 \\ C & 0.16 & 0.20 & 0.17 & 0.56 & 0.12 & 0.07 & 0.18 & 0.33 & -0.18 & -0.02 & 0.27 \\ \beta & 0.18 & 0.17 & 0.18 & 0.26 & 0.13 & 0.08 & 0.21 & 0.33 & 0.04 & 0.01 & 0.19 \\ w & 1.31 & 1.23 & 1.50 & 1.78 & 1.88 & 1.61 & 1.76 & 3.10 & 1.06 & 1.04 & 1.70 \\ \varphi & 3.04 & 3.80 & 5.23 & 1.83 & 2.29 & 5.34 & 0.92 & 2.55 & 6.17 & 7.02 & 6.20 \\ \hline \end{array} \]

\small \hskip 1 cm Table 1: parameters for the fits of the indices being considered (1987).

\vskip 0.2 cm

\normalsize

One can see that the majority of parameters are quite similar, and that all markets, with the exception of South Korea, which seems not to have felt the crisis at all, behaved in similar ways during and immediately after the crash. The similarity of the coefficients for the Dow Jones and the S\%P  500 indices, both from the NYSE, can be a guide to the precision of our fits. Parameter $A$ seems significant up to the second decimal digit; parameter $B$ is less accurate, with precision only up to the first decimal digit. The remaining paramaters all seem significant up to the second decimal digit.

The parameter $A$ shows that the Hong Kong, West Germany, and Mexico suffered the most severe losses after the crash. Parameter $B$ shows the strength of the initial shock to each market. Hong Kong and Brazil seem to have felt the strongest shocks. Parameter $\alpha $ shows how fast the shocks were absorbed. The initial shock fell less steeply for West Germany and the UK.

Parameters $B$ and $\beta $ show how fast the initial oscillations declined, and the initial amplitude of these oscillations. They were strongest for Hong Kong, but also diminished faster in that market.

One key parameter to be looked at is the frequency $w$ of the markets. They show that all markets, with the exception of Brazil, reacted with the same volatility just after the crash. This is a tendency that endured through the crises we are studying. The phase $\varphi $ just shows how long it took for each market to be struck by the severe after-crash oscillations. Since the South Korean market has such distinct parameters, we are not considering this market in our analysis.

\subsection{The crash of 1998 - Russian crisis}

Following the so called Asian crisis, in 1997, the prices of commodities fell worldwide. Russia, whose economy is heavily based on exports of commodities, was much affected by that. Other internal factors, like the war in Chechnya, drove the Russian economy to the brink of collapse. As many countries had money invested in Russia, the crisis spread to foreign markets and caused a crash from which they recovered slowly.

Figure 8 shows the Dow Jones index from February/1998 to February/1999, with the period that goes from from 08/25/1998 to 10/14/1998 highlighted, which is the time span we are considering for our model.



\vskip 0.8 cm

The graphs of the best fits that were obtained for the 1998 crash, from 08/25/1998 to 10/14/1998, are shown now. Here, we have enforced the restriction that $\beta \geq 0$. Had we not done so, then many of the oscillations would grow out of control, for the second oscillations after the crash were usually larger than the first ones.

\vskip 2.1 cm

%

\vskip 4 cm

The parameters are displayed together in the next table.

\small

\[ \begin{array}{|c||c|c|c|c|c|c|c|c|c|c|c|} \hline \text{Parameter} & \text{Dow Jones} & \text{S\&P 500} & \text{Nasdaq} & \text{Hang Seng} & \text{Nikkei} & \text{Dax} & \text{FTSE} & \text{Ibovespa} & \text{IPC} & \text{Kospi} & \text{Asx} \\ \hline A & 0.98 & 0.97 & 0.98 & 0.97 & 1.03 & 1.06 & 1.02 & 1.03 & 0.92 & 1.00 & 1.00 \\ B & 0.24 & 0.24 & 0.40 & 0.25 & 0.24 & 0.29 & 0.19 & 0.38 & 0.23 & 0.00 & 0.80 \\ \alpha  & 0.56 & 0.59 & 0.78 & 4.23 & 1.77 & 0.83 & 0.74 & 0.84 & 1.00 & 0.68 & 5.18 \\ C & 0.02 & 0.03 & 0.05 & 0.06 & 0.02 & 0.02 & 0.01 & 0.03 & 0.03 & 0.01 & 0.04 \\ \beta & 0.00 & 0.03 & 0.10 & 0.00 & 0.00 & 0.00 & 0.00 & 0.00 & 0.00 & 0.00 & 0.17 \\ w & 1.66 & 1.65 & 1.62 & 0.72 & 1.37 & 1.50 & 1.64 & 2.68 & 2.64 & 1.34 & 0.67 \\ \phi & 4.77 & 4.77 & 4.65 & 1.64 & 2.63 & 4.36 & 3.61 & 4.71 & 4.25 & 4.29 & 0.96 \\ \hline \end{array} \]

\small \hskip 1 cm Table 2: parameters for the fits of the indices being considered (1998).

\vskip 0.3 cm

\normalsize

The first striking feature is that most of the parameters $\beta $ are set to zero. This was built into the fitting process, for the oscillations just after the crash actually grew stronger in most markets, and that would ruin the approximation for later periods. Most countries showed faster recoveries than in 1987, what is shown by the larger values of parameter $A$. Brazil and Mexico had larger volatilities after the shock, followed by large drops in their indices. In the case of Brazil, the crisis was followed by strong speculative attacks against the local currency, and the market was very insecure at the time. Australia seemed not affected by the international financial crisis.

\subsection{The crash of 2001 - September $\mathbf{11}$}

In the morning of September 11, 2001, the USA suffered the most severe terrorist attack in all its history. The death toll was close to 3,000, including the terrorists who perpetrated the attack. The world watched in horror as two airplanes were shown live colliding with the twin towers of the World Trade Center, in New York. On that day, panic striken traders tried to get their money out of the stock exchanges and place their money in safer investments. Trade was cancelled in both the NYSE and the Nadaq. Markets recovered fast after the shock, for the economy had not been hit significantly and the stock market was healthy.

The S\&P 500 index from March/2001 to March/2001 is shown in figure 10. The period from 09/10/2001 to 10/22/2001, which we shall analyze, is highlighted.



\vskip 0.8 cm

We next show the graphs of the best fits that were obtained for the 2001 crash, from 09/10/2001 to 10/22/2001, together with the appropriate parameters. September 11 was a fine example of a crash triggered by some external news, without reasons for it of its own. In what follows, note that nearly all of the indices had a fast recovery after the first drop and the first two oscillations.

\vskip 1.8 cm

%

\vskip 3.7 cm

The parameters are displayed together in the next table.

\small

\[ \begin{array}{|c||c|c|c|c|c|c|c|c|c|c|c|} \hline \text{Parameter} & \text{Dow Jones} & \text{S\&P 500} & \text{Nasdaq} & \text{Hang Seng} & \text{Nikkei} & \text{Dax} & \text{FTSE} & \text{Ibovespa} & \text{IPC} & \text{Kospi} & \text{Asx} \\ \hline A & 0.94 & 0.94 & 0.93 & 0.95 & 0.97 & 0.93 & 0.43 & 0.98 & 0.95 & 0.50 & 0.96 \\ B & 0.45 & 0.32 & 0.30 & 0.18 & 0.12 & 6.44 & 0.57 & 0.92 & 0.39 & 0.56 & 0.14 \\ \alpha & 1.03 & 0.99 & 0.67 & 0.64 & 1.46 & 1.27 & 0.00 & 1.21 & 1.51 & 0.01 & 0.44 \\ C & 0.04 & 0.02 & 0.02 & -0.03 & 0.03 & -0.05 & -0.03 & 0.06 & 0.02 & -0.01 & 0.02 \\ \beta & 0.06 & 0.01 & 0.03 & 0.00 & 0.06 & 0.00 & 0.09 & 0.08 & 0.00 & -0.03 & 0.06 \\ w & 1.36 & 1.29 & 1.60 & 1.88 & 1.98 & 1.10 & 2.24 & 0.85 & 1.41 & 1.06 & 1.80 \\ \phi & 4.62 & 4.26 & 5.77 & 4.04 & 1.53 & 3.09 & 2.78 & -0.10 & 4.41 & 7.44 & -0.04 \\ \hline \end{array} \]

\small \hskip 1 cm Table 3: parameters for the fits of the indices being considered (2001).

\vskip 0.3 cm

\normalsize

What we have here is a much faster dissipating shock (higher values of $\alpha $ with small oscillations in its aftermath and small values of $B$). The most volatile markets after the shock were Japan and the UK (higher values for $w$). The low volatility for Brazil is mainly due to a choice between two values of $w$, one more volatile and the other less volatile, in the fitting process.

\subsection{The crash of 2008 - subprime mortgage Crisis}

\normalsize

We now analyze a somewhat different type of crash which occured recently, triggered by a subprime mortgage crisis in the USA. This crash is different because it is much slower than the others, with a mixture of the recipe we have seen in the other crashes. In order to describe this difference, let us consider the Dow Jones index from 09/12/2008 to 10/27/2008 (figure 12). The highlighted are shows a period where the index fell by a large amount, but one can easily notice the market was already going down before that.

\begin{pspicture}(-5,-0.5)(3.5,5.3)
\psset{xunit=0.03,yunit=0.006}
\psline{->}(1,1)(270,0) \psline{->}(1,0)(0,800) \rput(296,0){month} \rput(33,800){S\&P 500}\small \psline(1,-17)(1,17) \rput(1,-51){3/08} \psline(43,-17)(43,17) \rput(43,-51){5/08} \psline(85,-17)(85,17) \rput(85,-51){7/08} \psline(128,-17)(128,17) \rput(128,-51){9/08} \psline(172,-17)(172,17) \rput(172,-51){11/08} \psline(213,-17)(213,17) \rput(213,-51){01/09} \psline(252,-17)(252,17) \rput(252,-51){03/09} \psline(-3,0)(3,0) \rput(-12.5,0){$700$} \psline(-3,100)(3,100) \rput(-12.5,100){$800$} \psline(-3,200)(3,200) \rput(-12.5,200){$900$} \psline(-3,300)(3,300) \rput(-14,300){$1000$} \psline(-3,400)(3,400) \rput(-14,400){$1100$} \psline(-3,500)(3,500) \rput(-14,500){$1200$} \psline(-3,600)(3,600) \rput(-14,600){$1300$} \psline(-3,700)(3,700) \rput(-14,700){$1400$}
\pspolygon*[linecolor=lightgray](136,0)(136,800)(167,800)(167,0)
\psline[linecolor=blue](1,631.34)(2,626.75)(3,633.70)(4,604.34)(5,593.37)(6,573.37)(7,620.65)(8,608.77)(9,615.48)
(10,588.14)(11,576.60)(12,630.74)(13,598.42)(14,629.51)(15,649.88)(16,652.99)(17,641.13)(18,625.76)(19,615.22)(20,622.70)
(21,670.18)(22,667.53)(23,669.31)(24,670.40)(25,672.54)(26,665.54)(27,654.49)(28,660.55)(29,632.83)(30,628.32)(31,634.43)
(32,664.71)(33,665.56)(34,690.33)(35,688.17)(36,675.94)(37,679.93)(38,688.82)(39,697.84)(40,696.37)(41,690.94)(42,685.59)
(43,709.34)(44,713.90)(45,707.49)(46,718.26)(47,692.57)(48,697.68)(49,688.28)(50,703.58)(51,703.04)(52,708.66)(53,723.57)
(54,725.35)(55,726.63)(56,713.40)(57,690.71)(58,694.35)(59,675.93)(60,685.35)(61,690.84)(62,698.26)(63,700.38)(64,685.67)
(65,677.65)(66,677.20)(67,704.05)(68,660.68)(69,661.76)(70,658.44)(71,635.49)(72,639.87)(73,660.03)(74,660.14)(75,650.93)
(76,637.81)(77,642.83)(78,617.93)(79,618.00)(80,614.29)(81,621.97)(82,583.15)(83,578.38)(84,580.00)(85,584.91)(86,561.52)
(87,562.90)(88,552.31)(89,573.69)(90,544.68)(91,553.39)(92,539.49)(93,528.30)(94,514.91)(95,545.36)(96,560.31)
(97,560.68)(98,560.00)(99,577.00)(100,582.18)(101,552.54)(102,557.76)(103,534.37)(104,563.19)(105,584.26)(106,567.38)
(107,560.31)(108,549.01)(109,584.88)(110,589.18)(111,566.06)(112,596.31)(113,605.31)(114,589.59)(115,585.82)(116,592.93)
(117,598.19)(118,578.60)(119,566.68)(120,574.54)(121,577.72)(122,592.19)(123,566.84)(124,571.51)(125,581.66)(126,600.68)
(127,582.82)(128,577.57)(129,574.98)(130,536.82)(131,542.31)(132,567.79)(133,524.51)(134,532.04)(135,549.05)(136,551.69)
(137,492.69)(138,513.60)(139,456.39)(140,506.51)(141,555.07)(142,507.09)(143,488.22)(144,485.87)(145,509.18)(146,513.27)
(147,406.42)(148,464.74)(149,461.06)(150,414.28)(151,399.23)(152,356.89)(153,296.23)(154,284.94)(155,209.92)(156,199.22)
(157,303.35)(158,298.01)(159,207.84)(160,246.43)(161,240.55)(162,285.40)(163,255.05)(164,196.78)(165,208.11)(166,176.77)
(167,148.92)(168,240.51)(169,230.09)(170,254.09)(171,268.75)(172,266.30)(173,305.75)(174,252.77)(175,204.88)(176,230.99)
(177,219.21)(178,198.95)(179,152.30)(180,211.29)(181,173.29)(182,150.75)(183,159.12)(184,106.58)(185,52.44)(186,100.03)
(187,151.81)(188,157.39)(189,187.68)(190,196.24)(191,116.21)(192,148.81)(193,170.74)(194,145.22)(195,176.07)(196,209.70)
(197,188.67)(198,199.24)(199,173.59)(200,179.73)(201,168.57)(202,213.18)(203,204.42)(204,185.28)(205,187.88)(206,171.63)
(207,163.16)(208,165.02)(209,172.80)(210,169.42)(211,190.64)(212,203.25)(213,231.80)(214,227.45)(215,234.70)(216,206.65)
(217,209.73)(218,190.35)(219,170.26)(220,171.79)(221,142.62)(222,143.74)(223,150.12)(224,105.22)(225,140.24)(226,127.50)
(227,131.95)(228,136.57)(229,145.71)(230,174.09)(231,145.14)(232,125.88)(233,125.44)(234,138.51)(235,132.23)(236,145.85)
(237,168.60)(238,169.89)(239,127.16)(240,133.74)(241,135.19)(242,126.84)(243,89.17)(244,88.42)(245,78.94)(246,70.05)
(247,43.33)(248,73.14)(249,64.90)(250,52.83)(251,35.09)
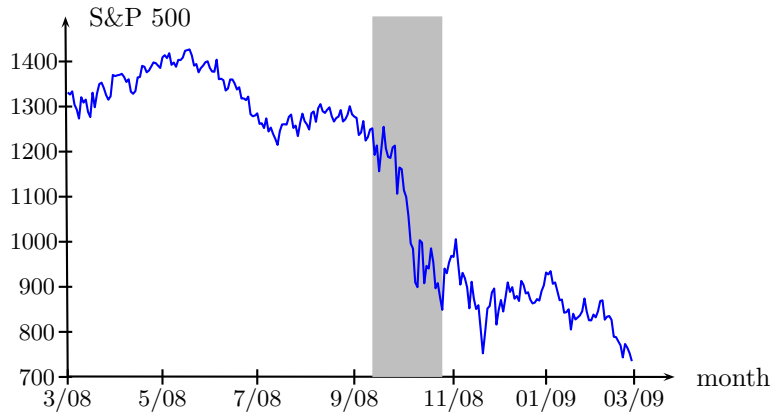
\rput(140,-140){\small Figure 12: S\&P 500 index from March/2008 to March/2008. The period from 09/12/2008 to 10/27/2008 is highlighted.}
\end{pspicture}

\vskip 0.8 cm

The next figure shows the graphs with the best fits for 2008, from 09/12/2008 to 10/27/2008.

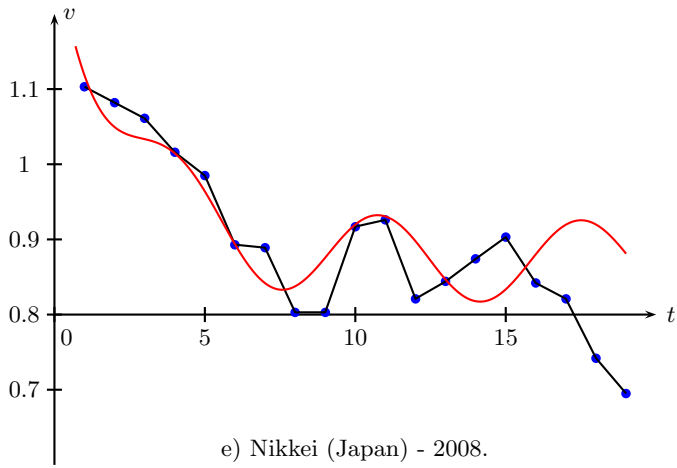
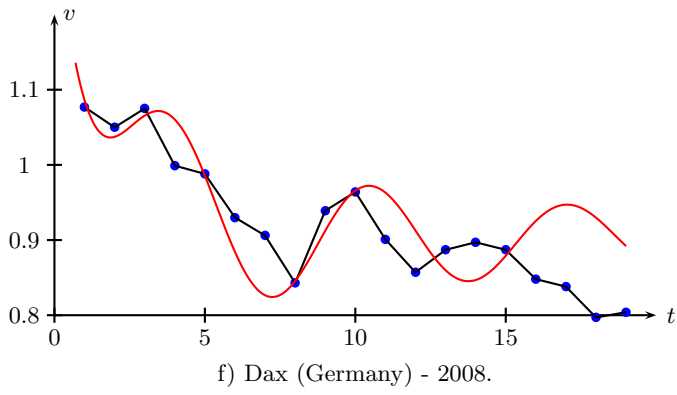
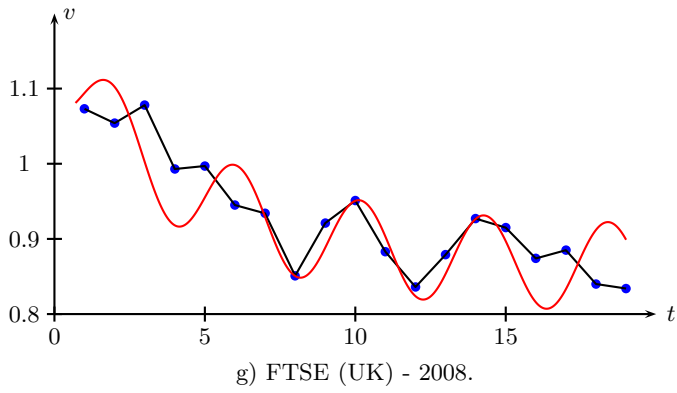
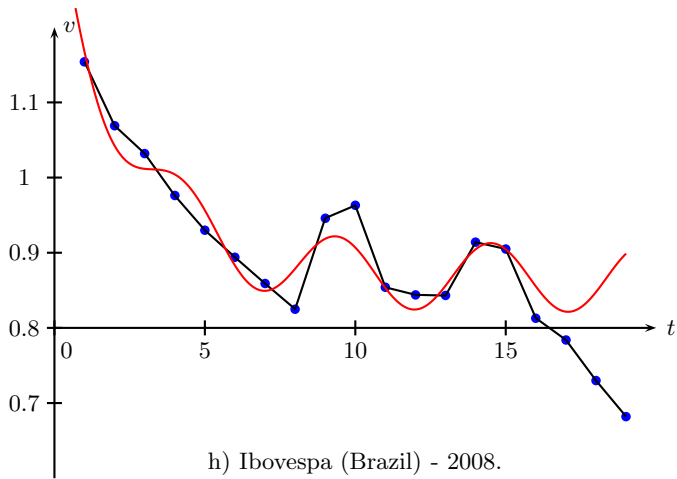
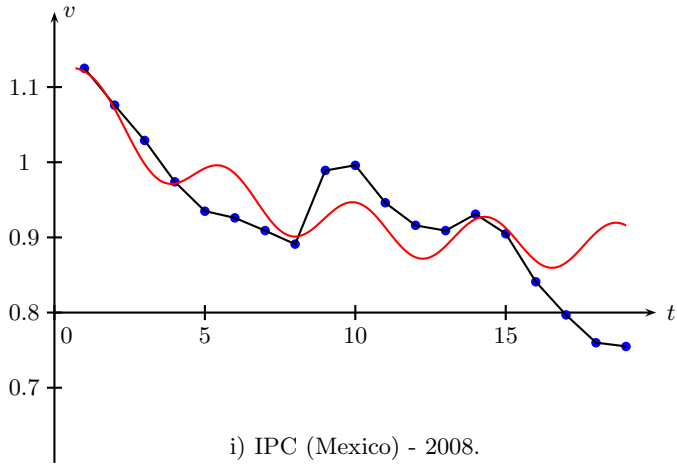
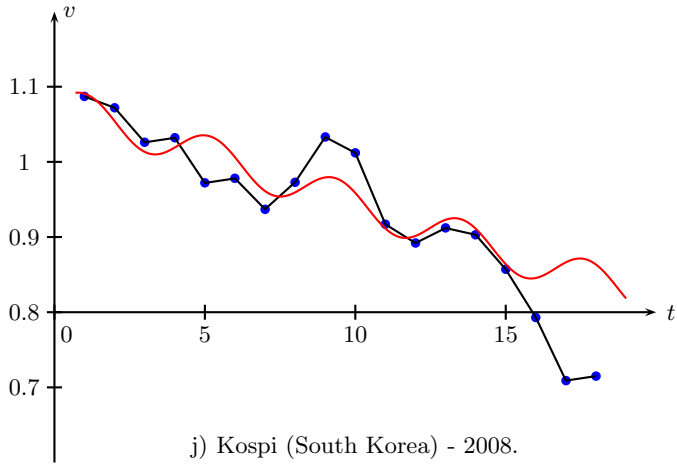
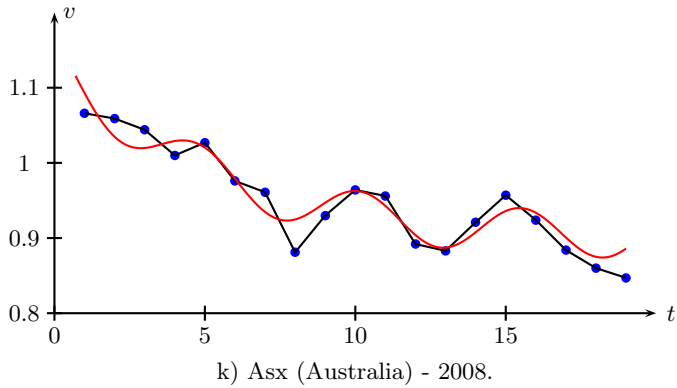
\begin{pspicture}(-1.5,0)(7.8,12)
\psset{xunit=0.4,yunit=10} \psline{->}(0,0.8)(20,0.8) \psline{->}(0,0.8)(0,1.20)
\psline(0,0.79)(0,0.81) \psline(5,0.79)(5,0.81) \psline(10,0.79)(10,0.81) \psline(15,0.79)(15,0.81) \small \rput(0,0.77){0} \rput(5,0.77){5} \rput(10,0.77){10} \rput(15,0.77){15} \psline(-0.25,0.8)(0.25,0.8) \psline(-0.25,0.9)(0.25,0.9) \psline(-0.25,1)(0.25,1) \psline(-0.25,1.1)(0.25,1.1) \rput(-1,0.8){$0.8$} \rput(-1,0.9){$0.9$} \rput(-1,1){$1$} \rput(-1,1.1){$1.1$}
\psdots[linecolor=blue](1,1.093)(2,1.058)(3,1.042)(4,1.005)(5,0.954)(6,0.935)(7,0.866)(8,0.853)(9,0.948)(10,0.940)
(11,0.866)(12,0.906)(13,0.894)(14,0.935)(15,0.912)(16,0.860)(17,0.877)(18,0.846)(19,0.825)
\psline(1,1.093)(2,1.058)(3,1.042)(4,1.005)(5,0.954)(6,0.935)(7,0.866)(8,0.853)(9,0.948)(10,0.940)
(11,0.866)(12,0.906)(13,0.894)(14,0.935)(15,0.912)(16,0.860)(17,0.877)(18,0.846)(19,0.825)
\psplot[linecolor=red,plotpoints=500]{0.7}{19}{2.72828182 x 0.5158 mul -1 mul exp 0.4998 mul 0.8926 add 1.1280 x mul 4.0526 sub 57.2958 mul cos 0.0575 mul 2.72828182 x 0.0130 mul -1 mul exp mul add}
\rput(20.5,0.8){$t$} \rput(0.5,1.2){$v$} \rput(10,0.72){\small Figure 13: a) Dow Jones - NYSE (USA) - 2008.}
\end{pspicture}

\begin{pspicture}(-1.5,0)(7.8,5.4)
\psset{xunit=0.4,yunit=10} \psline{->}(0,0.8)(20,0.8) \psline{->}(0,0.8)(0,1.20)
\psline(0,0.79)(0,0.81) \psline(5,0.79)(5,0.81) \psline(10,0.79)(10,0.81) \psline(15,0.79)(15,0.81) \small \rput(0,0.77){0} \rput(5,0.77){5} \rput(10,0.77){10} \rput(15,0.77){15} \psline(-0.25,0.8)(0.25,0.8) \psline(-0.25,0.9)(0.25,0.9) \psline(-0.25,1)(0.25,1) \psline(-0.25,1.1)(0.25,1.1) \rput(-1,0.8){$0.8$} \rput(-1,0.9){$0.9$} \rput(-1,1){$1$} \rput(-1,1.1){$1.1$}
\psdots[linecolor=blue](1,1.091)(2,1.048)(3,1.033)(4,0.994)(5,0.937)(6,0.926)(7,0.855)(8,0.845)(9,0.943)(10,0.938)
(11,0.853)(12,0.890)(13,0.884)(14,0.925)(15,0.898)(16,0.843)(17,0.854)(18,0.824)(19,0.798)
\psline(1,1.091)(2,1.048)(3,1.033)(4,0.994)(5,0.937)(6,0.926)(7,0.855)(8,0.845)(9,0.943)(10,0.938)
(11,0.853)(12,0.890)(13,0.884)(14,0.925)(15,0.898)(16,0.843)(17,0.854)(18,0.824)(19,0.798)
\psplot[linecolor=red,plotpoints=500]{0.7}{19}{2.72828182 x 0.5452 mul -1 mul exp 0.5159 mul 0.8845 add 1.1318 x mul 0.8675 sub 57.2958 mul cos -0.0612 mul 2.72828182 x 0.0114 mul -1 mul exp mul add}
\rput(20.5,0.8){$t$} \rput(0.5,1.2){$v$} \rput(10,0.72){\small b) S\&P 500 - NYSE (USA) - 2008.}
\end{pspicture}

\begin{pspicture}(-1.5,0)(7.8,5.4)
\psset{xunit=0.4,yunit=10} \psline{->}(0,0.8)(20,0.8) \psline{->}(0,0.8)(0,1.20)
\psline(0,0.79)(0,0.81) \psline(5,0.79)(5,0.81) \psline(10,0.79)(10,0.81) \psline(15,0.79)(15,0.81) \small \rput(0,0.77){0} \rput(5,0.77){5} \rput(10,0.77){10} \rput(15,0.77){15} \psline(-0.25,0.8)(0.25,0.8) \psline(-0.25,0.9)(0.25,0.9) \psline(-0.25,1)(0.25,1) \psline(-0.25,1.1)(0.25,1.1) \rput(-1,0.8){$0.8$} \rput(-1,0.9){$0.9$} \rput(-1,1){$1$} \rput(-1,1.1){$1.1$}
\psdots[linecolor=blue](1,1.082)(2,1.033)(3,1.018)(4,0.974)(5,0.917)(6,0.910)(7,0.860)(8,0.862)(9,0.964)(10,0.930)
(11,0.851)(12,0.898)(13,0.894)(14,0.925)(15,0.887)(16,0.844)(17,0.838)(18,0.811)(19,0.787)
\psline(1,1.082)(2,1.033)(3,1.018)(4,0.974)(5,0.917)(6,0.910)(7,0.860)(8,0.862)(9,0.964)(10,0.930)
(11,0.851)(12,0.898)(13,0.894)(14,0.925)(15,0.887)(16,0.844)(17,0.838)(18,0.811)(19,0.787)
\psplot[linecolor=red,plotpoints=500]{0.7}{19}{2.72828182 x 0.4710 mul -1 mul exp 0.4136 mul 0.8874 add 1.2362 x mul 4.8987 sub 57.2958 mul cos 0.0600 mul 2.72828182 x 0.0060 mul -1 mul exp mul add}
\rput(20.5,0.8){$t$} \rput(0.5,1.2){$v$} \rput(10,0.72){\small c) Nasdaq (USA) - 2008.}
\end{pspicture}

\begin{pspicture}(-1.5,5.4)(7.8,5.4)
\psset{xunit=0.4,yunit=10} \psline{->}(0,0.8)(20,0.8) \psline{->}(0,0.6)(0,1.20)
\psline(0,0.79)(0,0.81) \psline(5,0.79)(5,0.81) \psline(10,0.79)(10,0.81) \psline(15,0.79)(15,0.81) \small \rput(0.4,0.77){0} \rput(5,0.77){5} \rput(10,0.77){10} \rput(15,0.77){15} \psline(-0.25,0.7)(0.25,0.7) \psline(-0.25,0.8)(0.25,0.8) \psline(-0.25,0.9)(0.25,0.9) \psline(-0.25,1)(0.25,1) \psline(-0.25,1.1)(0.25,1.1) \rput(-1,0.7){$0.7$} \rput(-1,0.8){$0.8$} \rput(-1,0.9){$0.9$} \rput(-1,1){$1$} \rput(-1,1.1){$1.1$}
\psdots[linecolor=blue](1,1.073)(2,1.085)(3,1.053)(4,1.001)(5,1.001)(6,0.919)(7,0.950)(8,0.881)(9,0.972)(10,1.003)
(11,0.953)(12,0.907)(13,0.867)(14,0.913)(15,0.896)(16,0.850)(17,0.820)(18,0.752)(19,0.656)
\psline(1,1.073)(2,1.085)(3,1.053)(4,1.001)(5,1.001)(6,0.919)(7,0.950)(8,0.881)(9,0.972)(10,1.003)
(11,0.953)(12,0.907)(13,0.867)(14,0.913)(15,0.896)(16,0.850)(17,0.820)(18,0.752)(19,0.656)
\psplot[linecolor=red,plotpoints=500]{0.7}{19}{2.72828182 x 0.0136 mul -1 mul exp 0.8963 mul 0.1688 add 1.5125 x mul -0.4689 sub 57.2958 mul cos -0.0514 mul 2.72828182 x 0.0000 mul -1 mul exp mul add}
\rput(20.5,0.8){$t$} \rput(0.5,1.2){$v$} \rput(10,0.62){\small d) Hang Seng (Hong Kong) - 2008.}
\end{pspicture}

\newpage

{\white .}

\vskip 4.5 cm

\begin{pspicture}(-1.5,7)(7.8,7)
\psset{xunit=0.4,yunit=10} \psline{->}(0,0.8)(20,0.8) \psline{->}(0,0.6)(0,1.20)
\psline(0,0.79)(0,0.81) \psline(5,0.79)(5,0.81) \psline(10,0.79)(10,0.81) \psline(15,0.79)(15,0.81) \small \rput(0.4,0.77){0} \rput(5,0.77){5} \rput(10,0.77){10} \rput(15,0.77){15} \psline(-0.25,0.7)(0.25,0.7) \psline(-0.25,0.8)(0.25,0.8) \psline(-0.25,0.9)(0.25,0.9) \psline(-0.25,1)(0.25,1) \psline(-0.25,1.1)(0.25,1.1) \rput(-1,0.7){$0.7$} \rput(-1,0.8){$0.8$} \rput(-1,0.9){$0.9$} \rput(-1,1){$1$} \rput(-1,1.1){$1.1$}
\psdots[linecolor=blue](1,1.103)(2,1.082)(3,1.061)(4,1.016)(5,0.985)(6,0.893)(7,0.889)(8,0.803)(9,0.803)(10,0.917)
(11,0.926)(12,0.821)(13,0.844)(14,0.874)(15,0.903)(16,0.842)(17,0.821)(18,0.742)(19,0.695)
\psline(1,1.103)(2,1.082)(3,1.061)(4,1.016)(5,0.985)(6,0.893)(7,0.889)(8,0.803)(9,0.803)(10,0.917)
(11,0.926)(12,0.821)(13,0.844)(14,0.874)(15,0.903)(16,0.842)(17,0.821)(18,0.742)(19,0.695)
\psplot[linecolor=red,plotpoints=500]{0.7}{19}{2.72828182 x 0.4240 mul -1 mul exp 0.4656 mul 0.8712 add 0.9341 x mul 3.7923 sub 57.2958 mul cos 0.0600 mul 2.72828182 x 0.0060 mul -1 mul exp mul add}
\rput(20.5,0.8){$t$} \rput(0.5,1.2){$v$} \rput(10,0.62){\small e) Nikkei (Japan) - 2008.}
\end{pspicture}

\vskip 6.2 cm

\begin{pspicture}(-1.5,7)(7.8,7)
\psset{xunit=0.4,yunit=10} \psline{->}(0,0.8)(20,0.8) \psline{->}(0,0.8)(0,1.20)
\psline(0,0.79)(0,0.81) \psline(5,0.79)(5,0.81) \psline(10,0.79)(10,0.81) \psline(15,0.79)(15,0.81) \small \rput(0,0.77){0} \rput(5,0.77){5} \rput(10,0.77){10} \rput(15,0.77){15} \psline(-0.25,0.8)(0.25,0.8) \psline(-0.25,0.9)(0.25,0.9) \psline(-0.25,1)(0.25,1) \psline(-0.25,1.1)(0.25,1.1) \rput(-1,0.8){$0.8$} \rput(-1,0.9){$0.9$} \rput(-1,1){$1$} \rput(-1,1.1){$1.1$}
\psdots[linecolor=blue](1,1.077)(2,1.050)(3,1.075)(4,0.999)(5,0.988)(6,0.930)(7,0.906)(8,0.843)(9,0.939)(10,0.964)
(11,0.901)(12,0.857)(13,0.887)(14,0.897)(15,0.887)(16,0.848)(17,0.838)(18,0.797)(19,0.804)
\psline(1,1.077)(2,1.050)(3,1.075)(4,0.999)(5,0.988)(6,0.930)(7,0.906)(8,0.843)(9,0.939)(10,0.964)
(11,0.901)(12,0.857)(13,0.887)(14,0.897)(15,0.887)(16,0.848)(17,0.838)(18,0.797)(19,0.804)
\psplot[linecolor=red,plotpoints=500]{0.7}{19}{2.72828182 x 0.5763 mul -1 mul exp 0.5458 mul 0.9014 add 0.9558 x mul 0.6449 sub 57.2958 mul cos -0.1356 mul 2.72828182 x 0.0635 mul -1 mul exp mul add}
\rput(20.5,0.8){$t$} \rput(0.5,1.2){$v$} \rput(10,0.72){\small f) Dax (Germany) - 2008.}
\end{pspicture}

\vskip 5.3 cm

\begin{pspicture}(-1.5,7)(7.8,7)
\psset{xunit=0.4,yunit=10} \psline{->}(0,0.8)(20,0.8) \psline{->}(0,0.8)(0,1.20)
\psline(0,0.79)(0,0.81) \psline(5,0.79)(5,0.81) \psline(10,0.79)(10,0.81) \psline(15,0.79)(15,0.81) \small \rput(0,0.77){0} \rput(5,0.77){5} \rput(10,0.77){10} \rput(15,0.77){15} \psline(-0.25,0.8)(0.25,0.8) \psline(-0.25,0.9)(0.25,0.9) \psline(-0.25,1)(0.25,1) \psline(-0.25,1.1)(0.25,1.1) \rput(-1,0.8){$0.8$} \rput(-1,0.9){$0.9$} \rput(-1,1){$1$} \rput(-1,1.1){$1.1$}
\psdots[linecolor=blue](1,1.073)(2,1.054)(3,1.078)(4,0.993)(5,0.997)(6,0.945)(7,0.934)(8,0.851)(9,0.921)(10,0.951)
(11,0.883)(12,0.836)(13,0.879)(14,0.927)(15,0.915)(16,0.874)(17,0.885)(18,0.840)(19,0.834)
\psline(1,1.073)(2,1.054)(3,1.078)(4,0.993)(5,0.997)(6,0.945)(7,0.934)(8,0.851)(9,0.921)(10,0.951)
(11,0.883)(12,0.836)(13,0.879)(14,0.927)(15,0.915)(16,0.874)(17,0.885)(18,0.840)(19,0.834)
\psplot[linecolor=red,plotpoints=500]{0.7}{19}{2.72828182 x 0.2060 mul -1 mul exp 0.2790 mul 0.8570 add 1.5246 x mul 6.0690 sub 57.2958 mul cos -0.0615 mul 2.72828182 x 0.0021 mul -1 mul exp mul add}
\rput(20.5,0.8){$t$} \rput(0.5,1.2){$v$} \rput(10,0.72){\small g) FTSE (UK) - 2008.}
\end{pspicture}

\vskip 5.4 cm

\begin{pspicture}(-1.5,7)(7.8,7)
\psset{xunit=0.4,yunit=10} \psline{->}(0,0.8)(20,0.8) \psline{->}(0,0.6)(0,1.20)
\psline(0,0.79)(0,0.81) \psline(5,0.79)(5,0.81) \psline(10,0.79)(10,0.81) \psline(15,0.79)(15,0.81) \small \rput(0.4,0.77){0} \rput(5,0.77){5} \rput(10,0.77){10} \rput(15,0.77){15} \psline(-0.25,0.7)(0.25,0.7) \psline(-0.25,0.8)(0.25,0.8) \psline(-0.25,0.9)(0.25,0.9) \psline(-0.25,1)(0.25,1) \psline(-0.25,1.1)(0.25,1.1) \rput(-1,0.7){$0.7$} \rput(-1,0.8){$0.8$} \rput(-1,0.9){$0.9$} \rput(-1,1){$1$} \rput(-1,1.1){$1.1$}
\psdots[linecolor=blue](1,1.154)(2,1.069)(3,1.032)(4,0.976)(5,0.930)(6,0.894)(7,0.859)(8,0.825)(9,0.946)(10,0.963)
(11,0.854)(12,0.844)(13,0.843)(14,0.914)(15,0.905)(16,0.813)(17,0.784)(18,0.730)(19,0.682)
\psline(1,1.154)(2,1.069)(3,1.032)(4,0.976)(5,0.930)(6,0.894)(7,0.859)(8,0.825)(9,0.946)(10,0.963)
(11,0.854)(12,0.844)(13,0.843)(14,0.914)(15,0.905)(16,0.813)(17,0.784)(18,0.730)(19,0.682)
\psplot[linecolor=red,plotpoints=500]{0.7}{19}{2.72828182 x 0.4152 mul -1 mul exp 0.5020 mul 0.8663 add 1.2258 x mul 5.2150 sub 57.2958 mul cos 0.0453 mul 2.72828182 x 0.0000 mul -1 mul exp mul add}
\rput(20.5,0.8){$t$} \rput(0.5,1.2){$v$} \rput(10,0.62){\small h) Ibovespa (Brazil) - 2008.}
\end{pspicture}

\newpage

{\white .}

\vskip 4.5 cm

\begin{pspicture}(-1.5,7)(7.8,7)
\psset{xunit=0.4,yunit=10} \psline{->}(0,0.8)(20,0.8) \psline{->}(0,0.6)(0,1.20)
\psline(0,0.79)(0,0.81) \psline(5,0.79)(5,0.81) \psline(10,0.79)(10,0.81) \psline(15,0.79)(15,0.81) \small \rput(0.4,0.77){0} \rput(5,0.77){5} \rput(10,0.77){10} \rput(15,0.77){15} \psline(-0.25,0.7)(0.25,0.7) \psline(-0.25,0.8)(0.25,0.8) \psline(-0.25,0.9)(0.25,0.9) \psline(-0.25,1)(0.25,1) \psline(-0.25,1.1)(0.25,1.1) \rput(-1,0.7){$0.7$} \rput(-1,0.8){$0.8$} \rput(-1,0.9){$0.9$} \rput(-1,1){$1$} \rput(-1,1.1){$1.1$}
\psdots[linecolor=blue](1,1.125)(2,1.076)(3,1.029)(4,0.974)(5,0.935)(6,0.926)(7,0.909)(8,0.891)(9,0.989)(10,0.996)
(11,0.946)(12,0.916)(13,0.909)(14,0.931)(15,0.905)(16,0.841)(17,0.797)(18,0.760)(19,0.755)
\psline(1,1.125)(2,1.076)(3,1.029)(4,0.974)(5,0.935)(6,0.926)(7,0.909)(8,0.891)(9,0.989)(10,0.996)
(11,0.946)(12,0.916)(13,0.909)(14,0.931)(15,0.905)(16,0.841)(17,0.797)(18,0.760)(19,0.755)
\psplot[linecolor=red,plotpoints=500]{0.7}{19}{2.72828182 x 0.2070 mul -1 mul exp 0.2583 mul 0.8827 add 1.4487 x mul 8.2124 sub 57.2958 mul cos 0.0315 mul 2.72828182 x 0.0000 mul -1 mul exp mul add}
\rput(20.5,0.8){$t$} \rput(0.5,1.2){$v$} \rput(10,0.62){\small i) IPC (Mexico) - 2008.}
\end{pspicture}

\vskip 6.2 cm

\begin{pspicture}(-1.5,7)(7.8,7)
\psset{xunit=0.4,yunit=10} \psline{->}(0,0.8)(20,0.8) \psline{->}(0,0.6)(0,1.20)
\psline(0,0.79)(0,0.81) \psline(5,0.79)(5,0.81) \psline(10,0.79)(10,0.81) \psline(15,0.79)(15,0.81) \small \rput(0.4,0.77){0} \rput(5,0.77){5} \rput(10,0.77){10} \rput(15,0.77){15} \psline(-0.25,0.7)(0.25,0.7) \psline(-0.25,0.8)(0.25,0.8) \psline(-0.25,0.9)(0.25,0.9) \psline(-0.25,1)(0.25,1) \psline(-0.25,1.1)(0.25,1.1) \rput(-1,0.7){$0.7$} \rput(-1,0.8){$0.8$} \rput(-1,0.9){$0.9$} \rput(-1,1){$1$} \rput(-1,1.1){$1.1$}
\psdots[linecolor=blue](1,1.087)(2,1.072)(3,1.026)(4,1.032)(5,0.972)(6,0.978)(7,0.937)(8,0.973)(9,1.033)(10,1.012)
(11,0.917)(12,0.892)(13,0.912)(14,0.903)(15,0.857)(16,0.793)(17,0.709)(18,0.715)
\psline(1,1.087)(2,1.072)(3,1.026)(4,1.032)(5,0.972)(6,0.978)(7,0.937)(8,0.973)(9,1.033)(10,1.012)
(11,0.917)(12,0.892)(13,0.912)(14,0.903)(15,0.857)(16,0.793)(17,0.709)(18,0.715)
\psplot[linecolor=red,plotpoints=500]{0.7}{19}{2.72828182 x 0.0045 mul -1 mul exp 3.0536 mul -1.9743 add 1.5078 x mul -1.6023 sub 57.2958 mul cos -0.0252 mul 2.72828182 x 0.0000 mul -1 mul exp mul add}
\rput(20.5,0.8){$t$} \rput(0.5,1.2){$v$} \rput(10,0.62){\small j) Kospi (South Korea) - 2008.}
\end{pspicture}

\vskip 3.5 cm

\begin{pspicture}(-1.5,9.6)(7.8,9.6)
\psset{xunit=0.4,yunit=10} \psline{->}(0,0.8)(20,0.8) \psline{->}(0,0.8)(0,1.20)
\psline(0,0.79)(0,0.81) \psline(5,0.79)(5,0.81) \psline(10,0.79)(10,0.81) \psline(15,0.79)(15,0.81) \small \rput(0,0.77){0} \rput(5,0.77){5} \rput(10,0.77){10} \rput(15,0.77){15} \psline(-0.25,0.8)(0.25,0.8) \psline(-0.25,0.9)(0.25,0.9) \psline(-0.25,1)(0.25,1) \psline(-0.25,1.1)(0.25,1.1) \rput(-1,0.8){$0.8$} \rput(-1,0.9){$0.9$} \rput(-1,1){$1$} \rput(-1,1.1){$1.1$}
\psdots[linecolor=blue](1,1.066)(2,1.059)(3,1.044)(4,1.010)(5,1.027)(6,0.976)(7,0.961)(8,0.881)(9,0.930)(10,0.964)
(11,0.956)(12,0.892)(13,0.883)(14,0.921)(15,0.957)(16,0.924)(17,0.884)(18,0.860)(19,0.847)
\psline(1,1.066)(2,1.059)(3,1.044)(4,1.010)(5,1.027)(6,0.976)(7,0.961)(8,0.881)(9,0.930)(10,0.964)
(11,0.956)(12,0.892)(13,0.883)(14,0.921)(15,0.957)(16,0.924)(17,0.884)(18,0.860)(19,0.847)
\psplot[linecolor=red,plotpoints=500]{0.7}{19}{2.72828182 x 0.1939 mul -1 mul exp 0.2473 mul 0.8973 add 1.1732 x mul 2.4802 sub 57.2958 mul cos -0.0304 mul 2.72828182 x 0.0000 mul -1 mul exp mul add}
\rput(20.5,0.8){$t$} \rput(0.5,1.2){$v$} \rput(10,0.72){\small k) Asx (Australia) - 2008.}
\end{pspicture}

\vskip 2.9 cm

The parameters are displayed together in the next table.

\small

\[ \begin{array}{|c||c|c|c|c|c|c|c|c|c|c|c|} \hline \text{Parameter} & \text{Dow Jones} & \text{S\&P 500} & \text{Nasdaq} & \text{Hang Seng} & \text{Nikkei} & \text{Dax} & \text{FTSE} & \text{Ibovespa} & \text{IPC} & \text{Kospi} & \text{Asx} \\ \hline A & 0.89 & 0.88 & 0.89 & 0.17 & 0.87 & 0.90 & 0.86 & 0.87 & 0.88 & -1.97 & 0.90 \\ B & 0.50 & 0.52 & 0.41 & 0.90 & 0.46 & 0.54 & 0.28 & 0.50 & 0.26 & 3.05 & 0.25 \\ \alpha & 0.52 & 0.54 & 0.47 & 0.01 & 0.42 & 0.58 & 0.21 & 0.42 & 0.21 & 0.00 & 0.20 \\ C & 0.06 & -0.06 & 0.06 & -0.05 & 0.06 & -0.14 & -0.06 & 0.04 & 0.03 & -0.02 & -0.03 \\ \beta & 0.01 & 0.01 & 0.01 & 0.00 & 0.01 & 0.06 & 0.00 & 0.00 & 0.00 & 0.00 & 0.00
\\ w & 1.13 & 1.13 & 1.24 & 1.51 & 0.93 & 0.96 & 1.52 & 1.22 & 1.45 & 1.52 & 1.17 \\ \phi & 4.05 & 0.87 & 4.90 & -0.47 & 3.79 & 0.64 & 6.07 & 5.21 & 8.21 & -1.60 & 2.48 \\ \hline \end{array} \]

\small \hskip 1 cm Table 4: parameters for the fits of the indices being considered (2008).


\normalsize

\vskip 0.3 cm

Once more, it was necessary to fix $\beta $ as being positive only, what drove some of its values to zero. Note that the values of $A$ are low, what signals a tendency of the market to continue its way down after this shock. The post-shock oscillations are all small, as shown by the small values of the paramater $C$, and do not dissipate fast (parameter $\beta $). The volatility now is nearly the same for all markets, what may be a hint that the markets being studied are now more mature.

\subsection{Multiple shocks versus secondary effects}

\normalsize

An analysis of the data prior to the first day studied for the crisis of 2008, now from 09/12/2008 to 10/27/2008, reveals (figure 14) that the best fit for this sample of data is a damped harmonic oscillator with longer wavelength, which means a lower value for the frequency $w$.

\vskip 1.3 cm

\begin{pspicture}(-1.5,9.6)(7.8,9.6)
\psset{xunit=0.2,yunit=10} \psline{->}(0,0.8)(35,0.8) \psline{->}(0,0.8)(0,1.20)
\psline(0,0.79)(0,0.81) \psline(5,0.79)(5,0.81) \psline(10,0.79)(10,0.81) \psline(15,0.79)(15,0.81) \psline(20,0.79)(20,0.81) \psline(25,0.79)(25,0.81) \psline(30,0.79)(30,0.81) \small \rput(0,0.77){0} \rput(5,0.77){5} \rput(10,0.77){10} \rput(15,0.77){15} \rput(20,0.77){20} \rput(25,0.77){25} \rput(30,0.77){30} \psline(-0.5,0.8)(0.5,0.8) \psline(-0.5,0.9)(0.5,0.9) \psline(-0.5,1)(0.5,1) \psline(-0.5,1.1)(0.5,1.1) \rput(-2,0.8){$0.8$} \rput(-2,0.9){$0.9$} \rput(-2,1){$1$} \rput(-2,1.1){$1.1$}
\psdots[linecolor=blue](1,1.153)(2,1.102)(3,1.116)(4,1.071)(5,1.112)(6,1.150)(7,1.112)(8,1.096)(9,1.093)(10,1.113)
(11,1.125)(12,1.046)(13,1.095)(14,1.093)(15,1.058)(16,1.042)(17,1.005)(18,0.954)(19,0.935)(20,0.866)(21,0.853)
(22,0.948)(23,0.940)(24,0.866)(25,0.906)(26,0.894)(27,0.935)(28,0.912)(29,0.860)(30,0.877)(31,0.846)(32,0.825)
\psline(1,1.153)(2,1.102)(3,1.116)(4,1.071)(5,1.112)(6,1.150)(7,1.112)(8,1.096)(9,1.093)(10,1.113)
(11,1.125)(12,1.046)(13,1.095)(14,1.093)(15,1.058)(16,1.042)(17,1.005)(18,0.954)(19,0.935)(20,0.866)(21,0.853)
(22,0.948)(23,0.940)(24,0.866)(25,0.906)(26,0.894)(27,0.935)(28,0.912)(29,0.860)(30,0.877)(31,0.846)(32,0.825)
\psplot[linecolor=red,plotpoints=500]{0.5}{33}{2.72828182 x 0.0014 mul -1 mul exp 9.0164 mul -7.8265 add 0.3669 x mul 4.0134 sub 57.2958 mul cos 0.0567 mul 2.72828182 x 0.0136 mul -1 mul exp mul add}
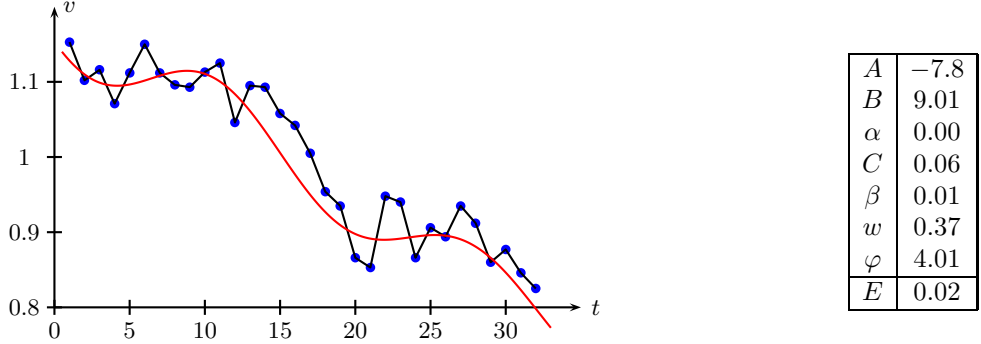
\rput(36,0.8){$t$} \rput(1,1.2){$v$} \rput(18,0.72){\small Figure 14: Dow Jones - NYSE (USA) - 2008. Fit with long wavelength.}
\end{pspicture}
\begin{minipage}{7 cm }
\[ \begin{array}{|c|c|} \hline A & -7.8 \\ B & 9.01 \\ \alpha & 0.00 \\ C & 0.06 \\ \beta & 0.01 \\ w & 0.37 \\ \varphi & 4.01 \\ \hline E & 0.02 \\ \hline \end{array} \]
\end{minipage}

\vskip 1.3 cm

This behavior is mimicked by all the other markets we are studying. Note, however, that this fit does not capture the minor oscillations within the wave described nor the larger fall that occurs at around day 14 (10/01/2008).

In order to try to make our model more adequate, we shall introduce a second shock, which occurs at time $t_0$, and that makes the Dow Jones index drop with greater intensity. Such a shock can be written as $\epsilon \e ^{-\zeta t}H(t-t_0)$, where $H(t-t_0)$ is the Heaviside function, given by
\begin{equation}
\label{heaviside}
H(t-t_0)=\left\{ \begin{array}{l} 0\ ,\ t<t_0,\\ 1\ ,\ t\geq t_0,\end{array} \right.
\end{equation}
so that the shock only begins at time $t_0$, which shall be considered a new parameter of the model, along with parameters $\epsilon $ and $\zeta $.

Now, from the point of view of damped harmonic oscillators, model (\ref{oscil}) becomes
\begin{equation}
\label{eqheavi}
m\ddot P+\gamma \dot P+kP=kP^*+\delta \e^{-\alpha t}+\epsilon \e^{-\zeta t}H(t-t_0)\ .
\end{equation}

This differential equation can be solved using the Laplace transform. The oscillatory solution is
\begin{equation}
\label{solheavi}
P(t)=A+B\e^{-\alpha t}+C\e^{-\beta t}\cos (wt-\varphi )+D\e^{-\zeta (t-t_0)}H(t-t_0)+E\e^{-\beta (t-t_0)}\cos \left( w(t-t_0)-\eta \right) H(t-t_0)\ ,
\end{equation}
where
\begin{eqnarray}
 & & A=P^*\ \ ,\ \ B=\frac{\delta }{m\alpha ^2-\gamma \alpha +k}\ \ ,\ \ \beta =\frac{\gamma }{2m}\ \ ,\ \ w=\frac{\sqrt{4mk-\gamma ^2}}{2m}\ ,\\
 & & \varphi =\arctg \left( \frac{s+1+(\gamma /m)(P_0-P^*)}{B}+\alpha -\beta \right) \ \ ,\ \ C=-\frac{B}{\cos \varphi }\ \ ,\\
  & & D=\frac{\epsilon }{m\zeta ^2-\gamma \zeta +k}\ \ ,\ \ \eta=-\arctg \left( \zeta -\beta \right) \ \ ,\ \ E=-\frac{D}{\cos \eta }\ .
\end{eqnarray}




Fitting the new parameters, as well as refitting the previous ones, we obtain the following curve.

\newpage

{\white .}

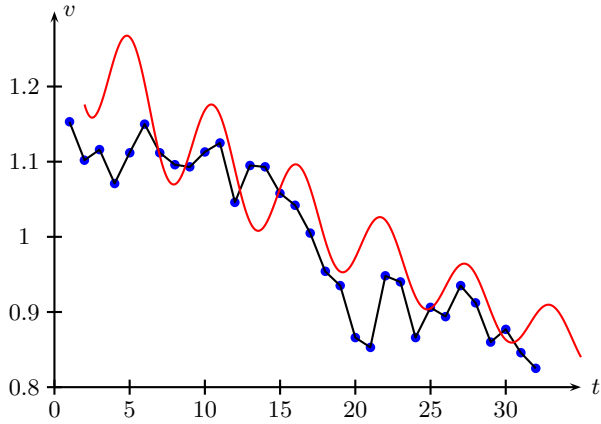
\begin{pspicture}(-1.5,10.5)(8.4,10.5)
\psset{xunit=0.2,yunit=10} \psline{->}(0,0.8)(35,0.8) \psline{->}(0,0.8)(0,1.3)
\psline(0,0.79)(0,0.81) \psline(5,0.79)(5,0.81) \psline(10,0.79)(10,0.81) \psline(15,0.79)(15,0.81) \psline(20,0.79)(20,0.81) \psline(25,0.79)(25,0.81) \psline(30,0.79)(30,0.81) \small \rput(0,0.77){0} \rput(5,0.77){5} \rput(10,0.77){10} \rput(15,0.77){15} \rput(20,0.77){20} \rput(25,0.77){25} \rput(30,0.77){30} \psline(-0.5,0.8)(0.5,0.8) \psline(-0.5,0.9)(0.5,0.9) \psline(-0.5,1)(0.5,1) \psline(-0.5,1.1)(0.5,1.1) \psline(-0.5,1.2)(0.5,1.2) \rput(-2,0.8){$0.8$} \rput(-2,0.9){$0.9$} \rput(-2,1){$1$} \rput(-2,1.1){$1.1$} \rput(-2,1.2){$1.2$}
\psdots[linecolor=blue](1,1.153)(2,1.102)(3,1.116)(4,1.071)(5,1.112)(6,1.150)(7,1.112)(8,1.096)(9,1.093)(10,1.113)
(11,1.125)(12,1.046)(13,1.095)(14,1.093)(15,1.058)(16,1.042)(17,1.005)(18,0.954)(19,0.935)(20,0.866)(21,0.853)
(22,0.948)(23,0.940)(24,0.866)(25,0.906)(26,0.894)(27,0.935)(28,0.912)(29,0.860)(30,0.877)(31,0.846)(32,0.825)
\psline(1,1.153)(2,1.102)(3,1.116)(4,1.071)(5,1.112)(6,1.150)(7,1.112)(8,1.096)(9,1.093)(10,1.113)
(11,1.125)(12,1.046)(13,1.095)(14,1.093)(15,1.058)(16,1.042)(17,1.005)(18,0.954)(19,0.935)(20,0.866)(21,0.853)
(22,0.948)(23,0.940)(24,0.866)(25,0.906)(26,0.894)(27,0.935)(28,0.912)(29,0.860)(30,0.877)(31,0.846)(32,0.825)
\psplot[linecolor=red,plotpoints=500]{2}{35}{2.72828182 x 0.0212 mul -1 mul exp 0.7676 mul 0.4942 add 1.1199 x mul 2.800 sub 57.2958 mul cos -0.1022 mul 2.72828182 x 0.0300 mul -1 mul exp mul add 2.72828182 x 1.1471 mul -1 mul exp 0.3240 mul add 1.1199 x mul 3.9529 sub 57.2958 mul cos 0.0347 mul 2.72828182 x 0.0300 mul -1 mul exp mul add}
\rput(36,0.8){$t$} \rput(1,1.3){$v$} \rput(18,0.72){\small Figure 15: Dow Jones - NYSE (USA) - 2008. Fit with two shocks.}
\end{pspicture}
\begin{minipage}{6 cm }
\[ \begin{array}{|c|c|} \hline A & 0.49 \\ B & 0.70 \\ \alpha & 0.02 \\ C & -0.10 \\ \beta & 0.03 \\ w & 1.12 \\ \varphi & 3.01 \\ \hline \end{array} \ \ \ \ \begin{array}{|c|c|} \hline D & 0.32 \\ E & 0.03 \\ \zeta & 1.15 \\ \eta & 3.95 \\ t_0 & 13.55 \\ \hline E & 0.08 \\ \hline \end{array} \]
\end{minipage}

\vskip 2.4 cm

One can see the fit is not very satisfying. We shall try, then, a different approach, using a second mode of vibration and trying to fit it on the previous data.

\vskip 1.3 cm

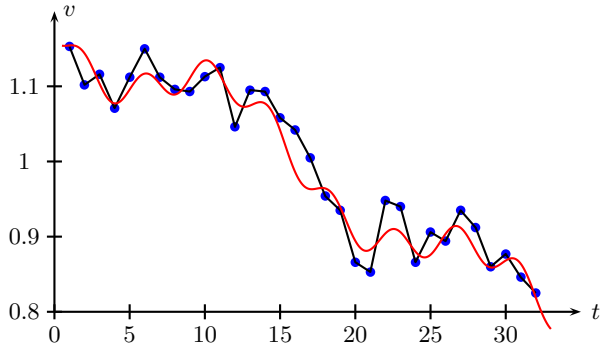
\begin{pspicture}(-1.5,9.6)(7.8,9.6)
\psset{xunit=0.2,yunit=10} \psline{->}(0,0.8)(35,0.8) \psline{->}(0,0.8)(0,1.20)
\psline(0,0.79)(0,0.81) \psline(5,0.79)(5,0.81) \psline(10,0.79)(10,0.81) \psline(15,0.79)(15,0.81) \psline(20,0.79)(20,0.81) \psline(25,0.79)(25,0.81) \psline(30,0.79)(30,0.81) \small \rput(0,0.77){0} \rput(5,0.77){5} \rput(10,0.77){10} \rput(15,0.77){15} \rput(20,0.77){20} \rput(25,0.77){25} \rput(30,0.77){30} \psline(-0.5,0.8)(0.5,0.8) \psline(-0.5,0.9)(0.5,0.9) \psline(-0.5,1)(0.5,1) \psline(-0.5,1.1)(0.5,1.1) \rput(-2,0.8){$0.8$} \rput(-2,0.9){$0.9$} \rput(-2,1){$1$} \rput(-2,1.1){$1.1$}
\psdots[linecolor=blue](1,1.153)(2,1.102)(3,1.116)(4,1.071)(5,1.112)(6,1.150)(7,1.112)(8,1.096)(9,1.093)(10,1.113)
(11,1.125)(12,1.046)(13,1.095)(14,1.093)(15,1.058)(16,1.042)(17,1.005)(18,0.954)(19,0.935)(20,0.866)(21,0.853)
(22,0.948)(23,0.940)(24,0.866)(25,0.906)(26,0.894)(27,0.935)(28,0.912)(29,0.860)(30,0.877)(31,0.846)(32,0.825)
\psline(1,1.153)(2,1.102)(3,1.116)(4,1.071)(5,1.112)(6,1.150)(7,1.112)(8,1.096)(9,1.093)(10,1.113)
(11,1.125)(12,1.046)(13,1.095)(14,1.093)(15,1.058)(16,1.042)(17,1.005)(18,0.954)(19,0.935)(20,0.866)(21,0.853)
(22,0.948)(23,0.940)(24,0.866)(25,0.906)(26,0.894)(27,0.935)(28,0.912)(29,0.860)(30,0.877)(31,0.846)(32,0.825)
\psplot[linecolor=red,plotpoints=500]{0.5}{33}{2.72828182 x 0.0014 mul -1 mul exp 9.0202 mul -7.8190 add 0.3649 x mul 4.3228 sub 57.2958 mul cos 0.0592 mul 2.72828182 x 0.0160 mul -1 mul exp mul add 1.5151 x mul 9.0778 sub 57.2958 mul cos 0.0201 mul 2.72828182 x 0.0000 mul -1 mul exp mul add}
\rput(36,0.8){$t$} \rput(1,1.2){$v$} \rput(18,0.72){\small Figure 16: Dow Jones - NYSE (USA) - 2008. Fit with two modes of vibration.}
\end{pspicture}
\begin{minipage}{7 cm }
\[ \begin{array}{|c|c|} \hline A & -7.82 \\ B & 9.02 \\ \alpha & 0.00 \\ C_1 & 0.06 \\ \beta_1 & 0.02 \\ w_1 & 0.36 \\ \varphi_1 & 4.32 \\ \hline \end{array} \ \ \ \ \begin{array}{|c|c|} \hline C_2 & 0.02 \\ \beta_2 & 0.00 \\ w_2 & 1.52 \\ \varphi_2 & 9.08 \\ \hline E & 0.02 \\ \hline \end{array} \]
\end{minipage}

\vskip 1.6 cm

The new model may be described by the following function:
\begin{equation}
P(t)=A+B\e^{-\alpha t}+C_1\e^{-\beta_1t}\cos (w_1t-\varphi_1)+C_2\e^{-\beta_2t}\cos (w_2t-\varphi_2)\ .
\end{equation}
Although it has less parameters than the multiple shocks model, it seems to describe best the behavior of this particular market. Such function is typical of coupled harmonic oscillators, resulting from the interaction of one or more markets and subject to a strong, rapidly dissipating shock. So, what this suggests is that the crisis of 2008 was most likely not the result of the response of individual markets to common, multiple external shocks, but might be the result of backreactions among markets.

We next make some comments on how financial markets are integrating in the past decades. This discussion is useful in our future attempts at linking the markets using an improvement of our model that acomodates many markets are coupled, damped harmonic oscillators.

\section{Correlation}

One good measure of how different markets are correlated during a period in time is the average of their correlation matrix. In the crisis of 1987, the average correlation between the indices being considered here during the period of crisis was $<C>=0.32$.

Another measure of how correlated different markets are is obtained by calculating the eigenvalues of the correlation matrix between the indices of those markets. The largest eigenvector corresponds to a ``market mode'', which expresses the common variation of the indices. In the case of the indices being considered for the crisis of 1987, the largest eigenvector of the correlation matrix between the many indices was $\lambda =4.89$. The graphs in figure 9 show  that some of the indices reacted much more strongly than others.

In the crisis of 1998, the average correlation between the indices being considered here during the period of crisis was $<C>=0.36$, and the largest eigenvector of the correlation matrix between the many indices was $\lambda =4.40$. The graphs in figure 11 show that few of the indices reacted much more strongly than others.

In the crisis of 2001, the average correlation between the indices being considered here during the period of crisis was $<C>=0.30$, and the largest eigenvector of the correlation matrix between the many indices was $\lambda =4.78$. Figure 13 shows good correlation between many of the indices.

In the crisis of 2008, the average correlation between the indices being considered here during the period of crisis was $<C>=0.46$, and the largest eigenvector of the correlation matrix between the many indices was $\lambda =7.71$. Figure 15 shows that all the indices considered reacted with the same amplitude, showing a strong dependance of each with the others.

\section{Mapping back}

We now try to map the coefficients we obtained back to the original coefficients, first of the damped harmonic oscillator model subject to an exponentially decreasing force, and then to the model of market dynamics with price expectations. Using equations (\ref{soldamped}) and (\ref{deltab}), we obtain the following relations between coefficients:
\begin{equation}
P^*=A\ \ ,\ \ c_1=C\ \ ,\ \ \gamma =2m\beta \ \ ,\ \ \delta =B(m\alpha ^2-\gamma \alpha +k)\ \ ,\ \ k=\frac{\gamma^2}{4m}+mw^2\ ,
\end{equation}
where $P^*$ would be the final value of the index if the market ever stabilized, $c_1$ may be related with the index at time zero by the equations
\begin{equation}
c_1=\frac{P^*-P_0-b}{\cos (-\varphi )}\ \ ,\ \ P_0=P(0)\ ,
\end{equation}
where $\delta $ is the initial strength of the shock, and $k$ is the constant of the spring, which sets how stiff the spring is. One can readily see that most equations depend on the same coefficient, $m$, which represents the mass of the system.

This leads us to a problem, that we dubbed ``the problem of mass'': what should represent the ``mass'' of a financial market? A massive market would be harder to be put into motion, but also slower to stop. It would also oscillate less rapidly than a market with little mass, and could be a major influence if markets would be linked in a network.

One of the possible answers would be that the mass of a market should be the amount of money it trades. We shall try this approach now, using as an example the data obtained for 2008. The following table contains the capitalization (in billions of dollars), of the 10 financial markets we are considering in our analysis at September, 2008, which is the time at which we do our analysis of that crisis.

\[ \begin{array}{|c|c|c|c|c|c|c|c|c|c|c|} \hline \text{Market} & \text{NYSE} & \text{Nasdaq} & \text{HK} & \text{Jap} & \text{Ger} & \text{UK} & \text{Bra} & \text{Mex} & \text{SK} & \text{Aus} \\ \hline \text{Capitalization} & 13,046 & 2,904 & 1,614 & 3,334 & 1,352 & 2,565 & 888 & 329 & 656 & 926 \\ \hline \end{array} \]

{\small \hskip 1.5 cm Table 5: market capitalizations, in billions of dollars, at September, 2008.}

\vskip 0.3 cm

\normalsize

Considering the NYSE as having mass one, and writing the masses of other markets in terms of it, we then obtain the following coefficients.

\[ \begin{array}{|c|c|c|c|c|c|c|c|c|c|c|c|} \hline \text{Market} & \text{DJ} & \text{S\&P} & \text{Nasdaq} & \text{HK} & \text{Jap} & \text{Ger} & \text{UK} & \text{Bra} & \text{Mex} & \text{SK} & \text{Aus} \\ \hline
m & 1.000 & 1.000 & 0.223 & 0.124 & 0.256 & 0.104 & 0.197 & 0.068 & 0.025 & 0.050 & 0.071 \\
P^* & 0.944 & 0.945 & 0.930 & 0.947 & 0.974 & 0.927 & 0.431 & 0.983 & 0.955 & 0.503 & 0.893 \\
c_1 & 0.039 & 0.025 & 0.025 & -0.027 & 0.033 & -0.053 & -0.033 & 0.058 & 0.024 & -0.010 & 0.273 \\
\gamma & 0.125 & 0.026 & 0.014 & 0.000 & 0.030 & 0.000 & 0.036 & 0.011 & 0.000 & -0.003 & 0.023 \\
k & 1.856 & 1.671 & 0.567 & 0.435 & 1.003 & 0.126 & 0.992 & 0.050 & 0.050 & 0.057 & 0.206 \\
\delta & 1.254 & 0.827 & 0.201 & 0.088 & 0.181 & 1.892 & 0.567 & 0.126 & 0.042 & 0.031 & 0.389 \\ \hline \end{array} \]

{\small \hskip 0.5 cm Table 6: coefficients of the damped harmonic oscillator model according to index (2008). Mass is chosen as capitalization.}

\vskip 0.3 cm

The results are not very clarifying, and this may result from the fact that the mass of the NYSE is much larger than most of the others, what leads to very small values for the dampening factor $\gamma $ and to the strength $\delta $ of the initial shocks to some markets.

Another approach to the problem of mass is to consider the inertial aspect of mass, which sets it as the difficulty in putting the market into motion. A variable that describes approximately how fast a market oscillates is its volatility, given by the absolute value of the standard deviation of its log-returns. Using the inverse of this parameter as mass, so that the mass of a market will be larger when volatility is small, and smaller when volatility is high, and setting this parameter as 1 for the NYSE during the period we are studying, we obtain the following table.

\[ \begin{array}{|c|c|c|c|c|c|c|c|c|c|c|c|} \hline \text{Market} & \text{DJ} & \text{S\&P} & \text{Nasdaq} & \text{HK} & \text{Jap} & \text{Ger} & \text{UK} & \text{Bra} & \text{Mex} & \text{SK} & \text{Aus} \\ \hline
m & 1.000 & 0.915 & 0.940 & 0.837 & 0.826 & 1.081 & 0.958 & 0.686 & 1.056 & 1.023 & 1.204 \\
P^* & 0.944 & 0.945 & 0.930 & 0.947 & 0.974 & 0.927 & 0.431 & 0.983 & 0.955 & 0.503 & 0.893 \\
c_1 & 0.039 & 0.025 & 0.025 & -0.027 & 0.033 & -0.053 & -0.033 & 0.058 & 0.024 & -0.010 & 0.273 \\
\gamma & 0.125 & 0.024 & 0.058 & 0.000 & 0.097 & 0.000 & 0.175 & 0.109 & 0.000 & -0.063 & 0.392 \\
k & 1.856 & 1.530 & 2.397 & 2.944 & 3.242 & 1.318 & 4.836 & 0.499 & 2.113 & 1.150 & 3.501 \\
\delta & 1.254 & 0.757 & 0.849 & 0.598 & 0.586 & 19.734 & 2.762 & 1.271 & 1.743 & 0.640 & 6.591
\\ \hline \end{array} \]

{\small Table 7: coefficients of the damped harmonic oscillator model according to index (2008). Mass is chosen as the inverse of the average volatility.}

\vskip 0.3 cm

The markets now have masses that are closer to one another, and the results are more similar. The problema with this definition of mass is that it leads to very different masses in 1987, though, influencing the remaining coefficients.

Another problem we have is that both definitions of mass change with time, and depend drastically on wether the stock markets are facing a crisis or not. We need a more stable definition for mass, and that is a topic for future research. That definition may come from the study of markets as a network of damped harmonic oscillators, and we hope we are able to reach it soon.

\section{Conclusion and future research}

Our work showed how stock markets may be modeled in periods of crisis as damped harmonic oscillators subject to an intense, but fast decreasing external force. The model may be put in analogy to a log-oscillator model for short time spans. The results show that some markets share some characteristics and differ in others like volatility after a crash, and absortion of the shocks generated by the crisis. It showed an increasing interaction between markets in times of crisis during the years via the correlation matrix of their log-returns. It also led us to believe that models of interacting markets as coupled harmonic oscillators may be a better way to understand the co-movements of financial markets in times of crisis. Our research shall concentrate on this topic from now on.

\vskip 0.6 cm

\noindent{\Large \bf Acknowledgements}

\vskip 0.4 cm

The authors thank for the support of this work by a grant from Insper, Instituto de Ensino e Pesquisa (L. Sandoval Jr.), and by a PIBIC grant from CNPq (I.P. Franca).


\begin{thebibliography}{99}


\bibitem{cont1} M. King and S. Wadhwani, {\sl Transmission of volatility between stock markets}, (1989) National Bureau of Economic Research working paper series, number 2910.

\bibitem{cont2} M. King, E. Sentana, and S. Wadhwani, {\sl Volatility and links between national stock markets}, (1990) National Bureau of Economic Research working paper series, number 3357.

\bibitem{cont3} J. Ammer and J. Mei, {\sl Measuring international economic linkages with stock market data}, (1993) Board of Governors of the Federal Reserve System, International finance discussion papers, number 449.

\bibitem{cont4} W-L Lin, R.F. Engle, and T. Ito, {\sl Do bulls and bears move across borders? International transmission of stock returns and volatility as the world turns}, Review of Financial Studies {\bf 7} (1994) 507-538.

\bibitem{cont5} C.B. Erb, C.R. Harvey, and T.E. Viskanta, {\sl Forecasting international equity correlations}, Financial Analyst Journal (November-December) (1994) 32-45.

\bibitem{cont6} T. Baig and I. Goldfajn, {\sl Financial market contagion in the Asian Crisis}, (1999) IMF Staff Papers {\bf 46}.

\bibitem{cont7} K. Forbes and R. Rigobon, {\sl No contagion, only interdependance: measuring stock market co-movements}, Journal of Finance {\bf 57} (2002) 2223-2261.

\bibitem{cont8} G. Corsetti, M. Pericoli, and M. Sbracia, {\sl Some contagion, some interdependance. More pitfalls in tests of financial contagion}, (2003).


\bibitem{time1} F. Longin and B. Solnik, {\sl Is the correlation in international equity returns constant: 1960-1990?}, J. of Int. Money and Finance {\bf 14} (1995) 3-26.

\bibitem{time2} G. Bekaert and C.R. Harvey, {\sl Time-varying world market integration}, The Journal of Finance {\bf 1} (1995) 403-444.

\bibitem{time3} G. de Santis and B. Gerard, {\sl International asset pricing and portfolio diversification with time-varying risk}, The Journal of Finance {\bf 52} (1997) 1881-1912.


\bibitem{vol1} B. Solnik, C. Boucrelle, and Y. Le Fur, {\sl International market correlation and volatility}, Financial Analysts Journal {\bf 52} (1996) 17-34.

\bibitem{vol2} I. Meric and G. Meric, {\sl Co-movements of European equity markets before and after the 1987 crash}, Multinational Finance Journal {\bf 1} (1997) 137-152.

\bibitem{vol3} F. Longin and B. Solnik, {\sl Correlation structure of international equity markets during extremely volatile periods}, Le Cashiers de Reserche, HEC Paris {\bf 646} (1999).

\bibitem{vol4} P. Hartmann, S. Straetmans, and C.G. de Vries, {\sl Asset market linkages in crisis periods}, (2001) Tinbergen Institute Discussion Paper, TI 2001-71/2.

\bibitem{vol5} F. Lillo, G. Bonanno, and R.N. Mantegna, {\sl Variety of stock returns in normal and extreme market days: the August 1998 crisis}, Proceedings of Empirical Science of Financial Fluctuations, Econophysics on the Horizon, Edited by H. Takayasu (2001).

\bibitem{vol6} A. Ang and J. Chen, {\sl Asymmetric correlations of equity portfolios}, Journal of Financial Economics {\bf 63} (2002) 443-494.

\bibitem{vol7} F. Longin and B. Solnik, {\sl Extreme correlation of international equity markets}, The Journal of Finance {\bf 56} (2001) 649-675.

\bibitem{vol8} I. Meric, S. Kim, J.H. Kim, and G. Meric, {\sl Co-movements of U.S., U.K., and Asian stock markets before and after September 11, 2001}, Journal of Money, Investiment and Banking {\bf 3} (2008) 47-57.

\bibitem{vol9} P. Cizeau, M. Potters, and J-P Bouchaud, {\sl Correlation structure of extreme stock returns}, Quantitative Finance {\bf 1} (2001) 217-222.

\bibitem{vol10} Y. Malevergne and D. Sornette, {\sl Investigating extreme dependances: concepts and tools}, Extreme Financial Risks (From dependance to risk management) (2006) Springer, Heidelberg.

\bibitem{vol11} R. Marshal and A. Zeevi, {\sl Beyond correlation: extreme co-movements between financial assets}, Working Paper, Columbia Business School (2002).

\bibitem{vol12} S.M. Bartram and Y-H Wang, {\sl Another look at the relationship between cross-market correlation and volatility}, Finance Research Letters {\bf 2} (2005) 75-88.

\bibitem{vol13} J. Knif, J. Kolari, and S. Pynnönen, {\sl What drives correlation between stock market returns?}, IMF Working Paper WP/07/157 (2007).


\bibitem{9801} A. Johansen and D. Sornette, {\sl Evidence of discrete scale invariance by canonical averaging}, Int. J. of Mod. Phys. C {\bf 9}, (1998) 433-447.

\bibitem{9802} A. Johansen and D. Sornette, {\sl Stock market crashes are outliers}, Eur. Phys. J. B {\bf 1}, (1998) 141-143.

\bibitem{9901} A. Johansen and D. Sornette, {\sl Critical crashes}, Risk {\bf 12}, (1999) 91-94.

\bibitem{9902} A. Johansen and D. Sornette, {\sl Fianancial ``anti-bubbles'': log-periodicity in gold and Nikkei collapses}, Int. J. Mod. Phys. C {\bf 10}, (1999) 563-575.

\bibitem{9903} A. Johansen and D. Sornette, {\sl Modeling the stock market prior to large crashes}, Eur. Phys. J. B {\bf 9}, (1999) 167-174.

\bibitem{9904} A. Johansen, O. Ledoit, and D. Sornette, {\sl Predicting financial crashes using discrete scale invariance}, J. of Risk {\bf 1}, (1999) 5-32.

\bibitem{0001} A. Johansen, O. Ledoit, and D. Sornette, {\sl Crashes as critical points}, Int. J. Mod. Theor. Applied Finance {\bf 3}, (2000).

\bibitem{0002} A. Johansen and D. Sornette, {\sl The nasdaq crash of April 2000: yet another example of log-periodicity in a speculative bubble ending in a crash}, Eur. Phys. J. B {\bf 17}, (2000) 319-328.

\bibitem{0003} A. Johansen and D. Sornette, {\sl Critical ruptures}, Eur. Phys. J. B {\bf 18}, (2000) 163-181.

\bibitem{0004} A. Johansen and D. Sornette, {\sl Evaluation of the quantitative prediction of a trend revearsal on the Japanese stock market in 1999}, Int. J. of Mod. Phys. C {\bf 11}, (2000) 359-364.

\bibitem{0005} A. Johansen, O. Ledoit, and D. Sornette, {\sl Crashes as critical points}, Int. J. of Theoretical and Applied Finance {\bf 3}, (2000) 219-255.

\bibitem{0101} A. Johansen and D. Sornette, {\sl Bubbles and anti-bubbles in Latin-American, Asian and Western stock markets: an empirical study}, Int. J. of Theoretical and Applied Finance {\bf 4} (6), (2001) 853-920.

\bibitem{0102} A. Johansen and D. Sornette, {\sl Finite-time singularity in the dynamics of the world population and economic indices}, Physica A {\bf 294}, (2001) 465-502.

\bibitem{0103} D. Sornette and A. Johansen, {\sl Significance of log-periodic precursors to financial crashes}, Quantitative Finance {\bf 1} (4), (2001) 452-471.

\bibitem{0201} A. Johansen and D. Sornette, {\sl Large stock market price drawdowns are outlier}, J. of Risk {\bf 4} (2), (2002) 69-110.

\bibitem{0202} C. Schulze, {\sl Sornette-Ide model for markets: trader expectations as imaginary part}, Int. J. Mod. Phys. C {\bf 14}, (2002).

\bibitem{0203} D. Sornette and W.X Zhou, {\sl The US 2000-2002 market descent: how much longer and deeper?}, Quantitative Finance {\bf 6}, (2002) 468-481.

\bibitem{0301} D. Sornette, {\sl Critical market crashes}, Phys. Rep. {\bf 378}, (2003) 1-98.

\bibitem{0302} W.X Zhou and D. Sornette, {\sl 2000-2003 real estate bubble in the UK but not in the USA}, Physica A {\bf 329}, (2003) 249-263.

\bibitem{0303} W.X Zhou and D. Sornette, {\sl Non-parametric analyses of log-periodic precursors to financial crashes}, Int. J. Mod. Phys. C {\bf 14}, (2003) 1107-1126.

\bibitem{0304} W.X. Zhou and D. Sornette, {\sl Evidence of a worldwide stock market log-periodic anti-bubble since mid-2000}, Physica A {\bf 330}, (2003) 543-583.

\bibitem{0305} D. Sornette, {\sl Why stock markets crash. Critical events in complex financial systems}, (2003) Princeton University Press.

\bibitem{0401} W.X. Zhou and D. Sornette, {\sl Antibubble and prediction of China's stock market and real-estate}, Physica A {\bf 337}, (2004) 243-268.

\bibitem{0402} J.V. Andersen and D. Sornette, {\sl Fearless versus fearful speculative financial bubbles}, Physica A {\bf 337}, (2004) 565-585.

\bibitem{0501} W.X. Zhou and D. Sornette, {\sl Testing the stability of the 2000 US stock maket ``antibubble''}, Physica A {\bf 348}, (2005) 428-452.

\bibitem{0502} M. Bartolozzi, S. Dro\`zd\`z, D.B. Leinweber, J. Speth, and A.W. Thomas, {\sl Self-similar log-periodic structures in western stock markets from 2000}, Int. J. Mod. Phys. C (2005) 1347-1361.

\bibitem{0601} A. Johansen and D. Sornette, {\sl Endogenous versus exogenous crashes in financial markets}, Cashiers Economiques de Bruxelles {\bf 49}, Special issue on nonlinear analysis (2006).

\bibitem{0801} L. Gazola, C. Fernandes, A. Pizzinge, and R. Riera, {\sl The log-periodic-AR(1)-GARCH(1,1) model for financial crashes}, The Eur. J. Phys. B {\bf 61}, (2008) 355-362.


\bibitem{leocorr} L. Sandoval Jr. and I. De P. Franca, {\sl Correlation of financial markets in times of crisis}, (2011) Physica A (in press).


\bibitem{ED1} R. Shone, {\sl Economic Dynamics: Phase Diagrams and their Economic Application}, second edition, (2003) Cambridge University Press.


\bibitem{B1} C. Hommes and T. Lux, {\sl Individual expectations and aggregate behavior in learning to forecast experiments}, (2008) Kiel Working Papers.


\bibitem{other01} K. Ideand and D. Sornette {\sl Oscillatory finite-time singularities in finance, population and rupture}, Physica A {\bf 307}, (2002) 63-106.

\bibitem{other02} D. Sornette, R. Woodard, and W.X. Zhou {\sl The 2006-2008 oil bubble and beyond}, Physica A {\bf 388}, (2009) 1571-1576.


\bibitem{new01} S. Dro\`zd\`z and P. O\'swi\c{e}cimka, {\sl World stock market: more sizeable trend revearsal likely in February/March 2010}, (2009) arXiv:0909.0418v3.

\bibitem{new02} Z.Q. Jiang, W.X. Zhou, D. Sornette, R. Woodard, K. Bastiaensen, and P. Cauwels, {\sl Bubble diagnosis and prediction of the 2005-2007 and 2008-2009 Chinese stock market bubbles}, Journal of Economic Behavior \& Organization, 74 (2010) 149–162.

\bibitem{new03} W. Yan, R. Woodard, and D. Sornette, {\sl Diagnosis and prediction of tipping points in financial markets: crashes and rebounds}, (2010) arXiv:1001.0265.

\bibitem{new04} V. Liberatore, {\sl Computational LPPL fit to financial bubbles}, (2010) arXiv:1003.2920.


\bibitem{crit01} A. Johansen, {\sl Comments on recent claims by Sornette and Zhou}, (2003) arXiv:con-mat/0302141.

\bibitem{crit02} D.S. Brée and N.L. Joseph, {\sl Fitting the log periodic power law to financial crashes: a crytical analysis}, (2010) arXiv:1002.1002v1.

\end{thebibliography}
\end{document}